\documentclass{aero2}
\usepackage[flushleft]{threeparttable}
\usepackage{longtable}
\usepackage{adjustbox}
\usepackage{booktabs,siunitx}
\usepackage{arydshln}
\usepackage{array}
\usepackage{xcolor}
\usepackage{multirow}
\usepackage{mathtools}

\ADsetup{
	title    = {Capacity of Sun-driven Lunar Swingby Sequences and Their Application in Asteroid Retrieval},
	author   = {Hongru Chen$^{1}$\cor{} \\
	},
	address  = {
		1. Department of Aeronautics and Astronautics, Kyushu University, Fukuoka, 8190395, Japan \sep
	},
	email    = {hongru.chen@hotmail.com},
	abstract = {
		For deep-space mission design, the gravity of the Sun and the Moon can be first considered and utilized. Their gravity can provide the energy change for launching spacecraft and retrieving spacecraft as well as asteroids. Regarding an asteroid retrieval mission, it can lead to the mitigation of asteroid hazards and an easy exploration and exploitation of the asteroid. 
		This paper discusses the application of the Sun-driven lunar swingby sequence for asteroid missions. Characterizing the capacity of this technique is not only interesting in terms of the dynamic insights but also non-trivial for trajectory design.  The capacity of a Sun-driven lunar swingby sequence is elucidated in this paper with the help of the ``Swingby-Jacobi'' graph. The capacity can be represented by a range of the Jacobi integral that encloses around 660 asteroids currently cataloged. To facilitate trajectory design, a database of Sun-perturbed Moon-to-Moon transfers, including multi-revolution cases is generated and employed. Massive trajectory options for spacecraft launch and asteroid capture can then be explored and optimized. Finally, a number of asteroid flyby, rendezvous, sample-return, and retrieval mission options enabled by the proposed technique are obtained.
	},
	keywords = {
		Asteroid mining \sep Planetary defense \sep Gravity assists \sep Moon-to-Moon transfers \sep Reachable  sets \sep Graphical methods\\
		
	},
}

\begin{document}
	
	\maketitle  
	
	
	\Nomenclature
	
\begin{center}
	\begin{tabular}{|P{.45\linewidth}|P{.45\linewidth}|}
		\hline
		NRHO & near-rectilinear halo orbit\\\hline
		SPMT & Sun-perturbed Moon-to-Moon transfer\\\hline
		GTO & geostationary transfer orbit\\\hline
		SE & Sun-Earth\\\hline
		CR3BP & circular-restricted three-body problem\\\hline
		GNC & guidance, navigation, and control operations\\\hline
		$\Delta v$ & impulsive velocity increment\\\hline
		$\bf{v}_{\infty}$, $v_{\infty}$ & vector and magnitude of the hyperbolic excess velocity with respect to a celestial body\\\hline
		$_\text{M}$, $_\text{E}$ & subscripts indicating references to the Moon and the Earth, respectively\\\hline
		$\varphi$ & swingby pump angle \\\hline
		$\delta$ & declination of the escape or incident trajectory \\\hline
		$C_3$ & characteristic energy (also $v^2_{\infty}$) with respect to Earth \\\hline
		$J$ & Jacobi integral in the SE-CR3BP \\\hline
		$ToF$ & time of flight\\\hline
		$\mathcal{A}$ & reachable set with a Sun-driven lunar swingby sequence\\\hline
		$\mathcal{A}_\text{p}$ & reachable set with a planar swingby sequence \\\hline
		$\mathcal{B}$ & reachable domain of $v_{\infty\text{E}}$ and $\delta$ \\\hline

		\end{tabular}
	\end{center}

	\section{Introduction}
	In space mission and trajectory design, gravity assists and low-energy transfers that take advantage of the gravity of celestial bodies are often employed to reduce fuel consumption as well as increase the mass of payloads. For all missions that depart from Earth, the Moon is the closest celestial body that can provide a gravity assist. In fact, in the Sun-Earth-Moon system a spacecraft is influenced by the joint gravitational effect from the Sun, the Earth, and the Moon. Trajectory design for missions such as the HITEN, ARTEMIS-THEMIS, NOZOMI, DESTINY, EQUULEUS, and Lunar Flashlight have employed lunar swingbys driven by solar tides~\cite{UESUGI1991347, FOLTA2012237, KAWAGUCHI2003189, Chen2015, Chen2016, GarciaYarnoz2016, LunarFlashlight2020}. The joint gravitational effect can assist the launch of a spacecraft and the capture of the returning spacecraft as well as samples from the visited body. However, the joint gravitational effect has not yet been well characterized for mission analysis and trajectory design. In this paper, the capacity of a Sun-driven lunar swingby sequence and its application for asteroid retrieval will be clarified.
	
	Asteroids are considered to be promising bonanza that may contain water and rare metals. These resources can be used to sustain astronauts’ lives, fuel spacecraft as well as power space bases. Developments on Earth can also benefit from asteroid mining. Furthermore, asteroids are of great scientific interest as they preserve pristine relics of the early solar system. Missions such as DAWN, NEAR, Hayabusa-1 and -2, and Osiris-REx have returned significant information and even samples that led to a better understanding of asteroids and the early solar system. An increasing number of missions, such as LUCY, PSYCHE, and MMX, will be visiting asteroids and asteroid-like small bodies. However, near-Earth asteroids (NEA) can be threats to Earth due to the constant possibility of Earth impacts. The Chelyabinsk meteor should be considered a warning. It is time to ``look up'' and tackle potential asteroid attacks. NASA and ESA are working on the DART and HERA missions to demonstrate asteroid deflection technologies~\cite{CHENG2018104, Michel_2022}. In this context, the concept of asteroid retrieval has also been proposed. Asteroid retrieval has the goal of bringing an asteroid or a part of it into the vicinity of the Earth. That will not only mitigate the threat but also allows for easy and constant visits to the asteroid. Moreover, the technology necessary for retrieving asteroids becomes more realistic. For instance, retrieving an entire small NEA – with a diameter of approximately 7 m and a mass on the order of 500,000 kg is believed to be possible by 2025~\cite{Brophy2012}.

	Asteroid retrieval was investigated from the perspectives of propulsion technologies and trajectory design in several studies. In this paper, the problem is discussed from the perspective of trajectory design. In this frame, the Sun-Earth libration-point orbits have been proposed for capturing asteroids~\cite{Baoyin_2010,Sanchez2012,LLADO2014176,Sanchez2016}. As this method mostly relies on natural capture by the gravity of the Sun and the Earth, it takes considerable time or $\Delta v$ to retrieve a target. In addition, libration-point orbits are not stable. Constant station-keeping maneuvers are needed, which is not feasible in the long run. Without stationkeeping, the captured asteroid could become a serious hazard to Earth. A lunar swingby can change the orbital energy more effectively. Landau et al. (2013)~\cite{Landau2013} and Gong and Li (2015)~\cite{Gong2015} proposed using lunar swingbys. Using this method, the $v_\infty$ with respect to the Moon is considered invariable,  and resulting orbits with their apogees beyond the Moon’s orbit will be constantly influenced by lunisolar perturbations, which are unstable and threatening to Earth. For stable capture, not only does the characteristic energy with respect to Earth have to be minimized, but also the $v_\infty$ with respect to the Moon.
	Urrutxua et al. (2015)~\cite{Urrutxua2015} included the gravity of the Moon in the Sun-Earth Hill's problem, and ran extensive simulations to reveal temporary capture trajectories prolonged by tens to hundreds of meters per second of $\Delta v$. Nevertheless, a concrete characterization of easily capturable orbit types (or objects) under the gravity of the Sun, the Earth, and the Moon, is still lacking. 
	
	
	The Sun-driven lunar swingby sequence investigated in this paper can change the $v_{\infty}$ with respect to both the Earth and the Moon, and thus can be used for launching spacecraft and capturing asteroids into a high lunar orbit or the Earth-Moon near-rectilinear halo orbit (NRHO), which is stable to the Earth. In this paper, the retrieval mission is considered in three phases. In the vicinity of the Earth, for the escape phase, a Sun-driven lunar swingby sequence is utilized to increase the $v_{\infty}$ of the spacecraft with respect to Earth, as well as target a certain direction for the asteroid. For the capture phase, a Sun-driven lunar swingby sequence is used to reduce the $v_{\infty}$ of the asteroid with respect to the Moon to an acceptably low level, which allows for an inexpensive insertion into orbits stably confined to the Moon. In the heliocentric transfer phase, the spacecraft rendezvous with the targeted asteroid, and the asteroid is sent back to Earth with permissible $\Delta v$. 
	There are around 15000 NEA currently cataloged. 
	Searching and optimizing possible heliocentric transfer and swingby sequences each time the asteroid database is updated is computationally expensive and time-consuming.  
	It is often desired to characterize the capacity (i.e., referred to as the reachable set, attainable set, or attractive set in the literature) of certain trajectory design methods for the purposes of obtaining initial guesses, pruning the search space, etc. (e.g., Ref.~\cite{Mingotti2011,Chilan2014,Bando2016,TOPPUTO20214193,Wu2023}). 
	Regarding the problem of interest in this paper, characterizing the capacity of Sun-driven lunar swingby sequences for reaching and capturing heliocentric objects is not only interesting in terms of dynamic insights but also non-trivial for mission planning. 
	
	Graphical methods are sometimes helpful for trajectory design and analysis. Strange and Longuski (2002)~\cite{Strange2002} proposed the Tisserand graph with $v_{\infty}$ contours for analyzing and planning gravity-assist sequences. It is based on the patched two-body approximation. The Poincaré map is useful in analyzing the three-body problem. Campagnola and Russell (2010)~\cite{Campagnola2010} further developed the Tisserand-Poincaré graph with the Tisserand (or Jacobi) contours that consider the third-body perturbation during a swingby. Ross and Scheeres (2007)~\cite{Ross2007} proposed the periapsis-Poincaré map for analyzing low-energy capture and escape. 
	Those graphs are applicable for star-planets and planet-moons systems. For the star-planet-moon system concerned in this work, the Swingby-Jacobi graph developed by Chen et al. (2016)~\cite{Chen2016} and Chen (2017)~\cite{Chen2017} is applicable. In addition to the Jacobi integral in the Sun-Earth system, the effect of lunar swingbys can also be portrayed. The double Tisserand graph recently developed by Martens and Bucci (2022)~\cite{Martens2022} is notable and also applicable for the star-planet-moon system. The double Tisserand graph is suitable for analyzing transfers to the moon, while the Swingby-Jacobi graph is suitable for analyzing transfers to the heliocentric region. The Swingby-Jacobi graph is adopted in this paper for elucidating the accessibility of a Sun-driven lunar swingby sequence to the heliocentric region. 
	
	This paper is organized as follows. In Sec.~\ref{Sec:Intro}, the dynamic model used for trajectory design and the mechanism of the sequence of Sun-perturbed Moon-to-Moon transfers (SPMT) and lunar swingbys for enhancing mission design are presented. The  Swingby-Jacobi graph is also introduced to visualize the effects of Sun-perturbed transfers and lunar swingbys. 
	Sec.~\ref{Sec:capacity} elucidates the capacity of a Sun-driven lunar swingby sequence with the help of the Swingby-Jacobi graph. It is characterized by an accessible range of the Jacobi integral and an accessible domain of the magnitude and declination of the $\bf{v}_{\infty}$ with respect to Earth. These ranges effectively exclude the definitely inaccessible asteroids and inaccessible heliocentric transfer options from among the optimized. The computation routine for optimal heliocentric transfers is presented in Sec.~\ref{Sec:helio}. 
	To facilitate patching SPMT segments and lunar swingbys for meeting the escape or capture conditions, a database of SPMT solutions is established. The effort of computing the database is described in Sec.~\ref{sec:database}. With the characterized capacity of this technique and the trajectory design tools, lists of possible asteroid flyby, rendezvous, sample-return, and retrieval missions for the next 20 years are obtained, which are presented in Sec.~\ref{Sec:results}. 
	Disadvantages and advantages of different Moon-to-Moon transfer types, namely the short transfer, Sun-perturbed transfer, and three-dimensional transfer, are discussed in Sec.~\ref{Sec:discussions}. Conclusions are given in the last section.
	
	\section{Dynamic Model and Swingby-Jacobi Graph}~\label{Sec:Intro}
	\subsection{Dynamic Model}
\begin{figure}[h!t]
		\begin{center}
			\includegraphics[width=0.7\hsize]{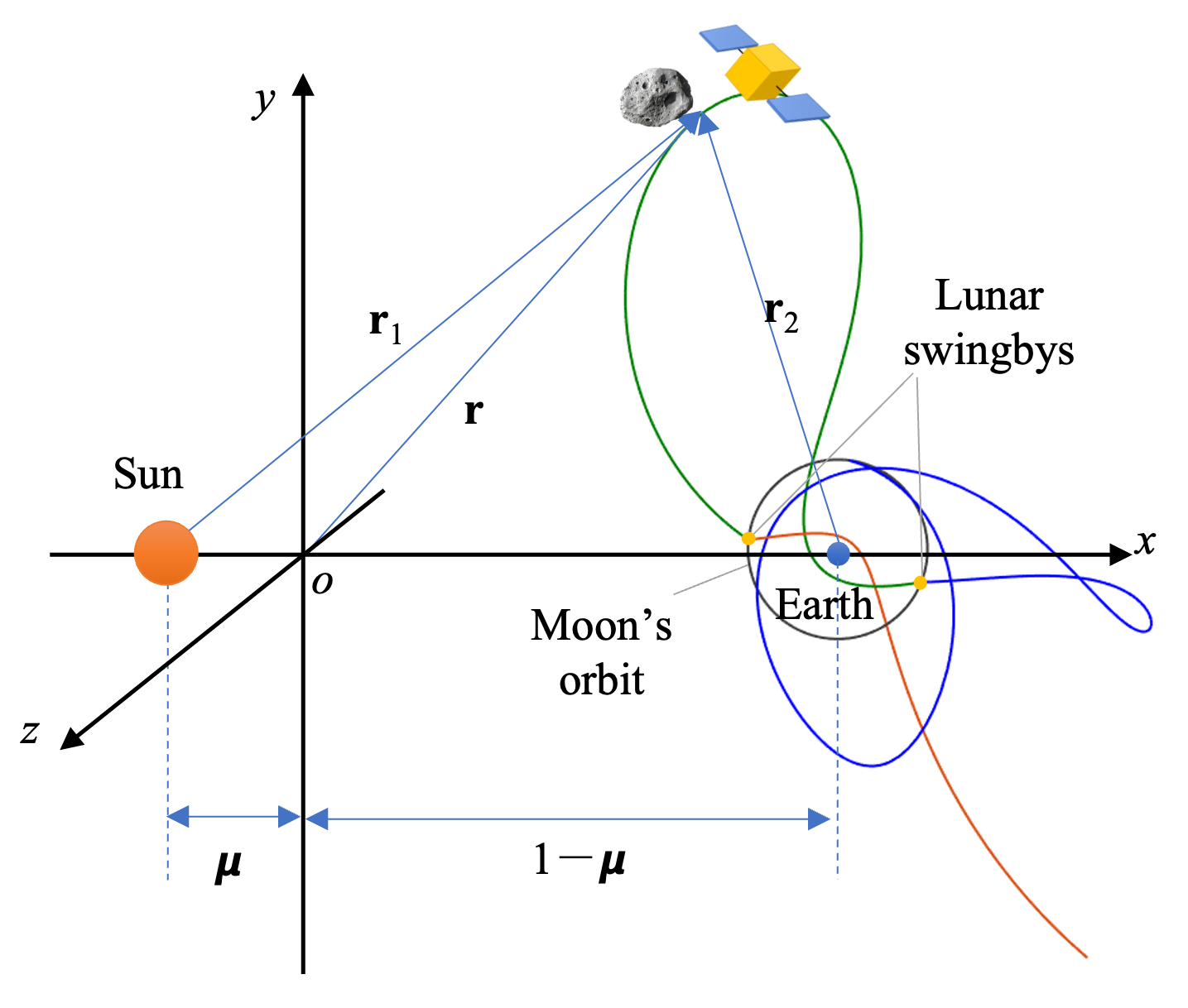}
			\caption{Schematic of the patched Sun-Earth CR3BP and lunar swingby model.}
			\label{fig:Model}
		\end{center}
	\end{figure}

	The motions of spacecraft and asteroid samples in the vicinity of the Earth are
	influenced by the gravity of the Sun, the Earth, and the Moon. To facilitate analysis and orbit design, the patched Sun-Earth circular restricted three-body problem (SE-CR3PB) and lunar swingby model is adopted. The synodic coordinate system, in which the Sun-Earth barycenter is put at the origin and the Sun and Earth are fixed on the x-axis, and the rotation direction is aligned with the z-axis, as shown in Fig.~\ref{fig:Model}, is used. Let $\mu$ denote the ratio of the mass of the Earth to the total mass of the two bodies, the distance between the Sun and the Earth, AU, be the length unit, and the time unit is set in the way that the period of the Earth's becomes $2\pi$. Then, the equations of motion in the SE-CR3BP are expressed as:
	\begin{align}\label{eq:EoM}
		\ddot x- 2\dot y &= {\partial U}/{\partial x}  \\
		\ddot y + 2\dot x & = {\partial U}/{\partial y} \\ 
		\ddot z&={\partial U}/{\partial z}
	\end{align}
	where $U$ denotes the pseudo-gravitational potential expressed as:
	\begin{equation}
		U=(x^2+y^2)/2+(1-\mu)/r_1+\mu/r_2
	\end{equation}
	where $\mu =\SI{3.0035e-6}{}$ for the Sun-Earth system, and the distances are:
	\begin{equation}
		r_1=\sqrt{(x+\mu)^2+y^2+z^2},~~~~~ r_2=\sqrt{(x-1+\mu)^2+y^2+z^2}
	\end{equation}
	An energy quantity, known as the Jacobi integral, holds in the CR3BP system. The Jacobi integral $J$ is expressed as: 
	\begin{equation}
		J=({{\dot{x}}^{2}}+{{\dot{y}}^{2}}+{{\dot{z}}^{2}}) - 2U
	\end{equation}
	
	Trajectories are considered in the SE-CR3BP most of the time. The lunar gravity is approximated by the lunar swingby, which is only effective at a lunar encounter. In other words, the object receives an instantaneous velocity change (i.e., $\bf{v}_{\infty}$ with respect to the Moon gets an instantaneous bend) at a lunar encounter. The inclination and eccentricity of the Moon's orbit are small and are set to zero in this simplified model. Such a SE-CR3BP plus lunar swingby model is schematically depicted in Fig.~\ref{fig:Model}.
	
	\subsection{Sun-perturbed Moon-to-Moon transfers}
\begin{figure}[h!t]
			\begin{center}
				\includegraphics[width=0.7\hsize]{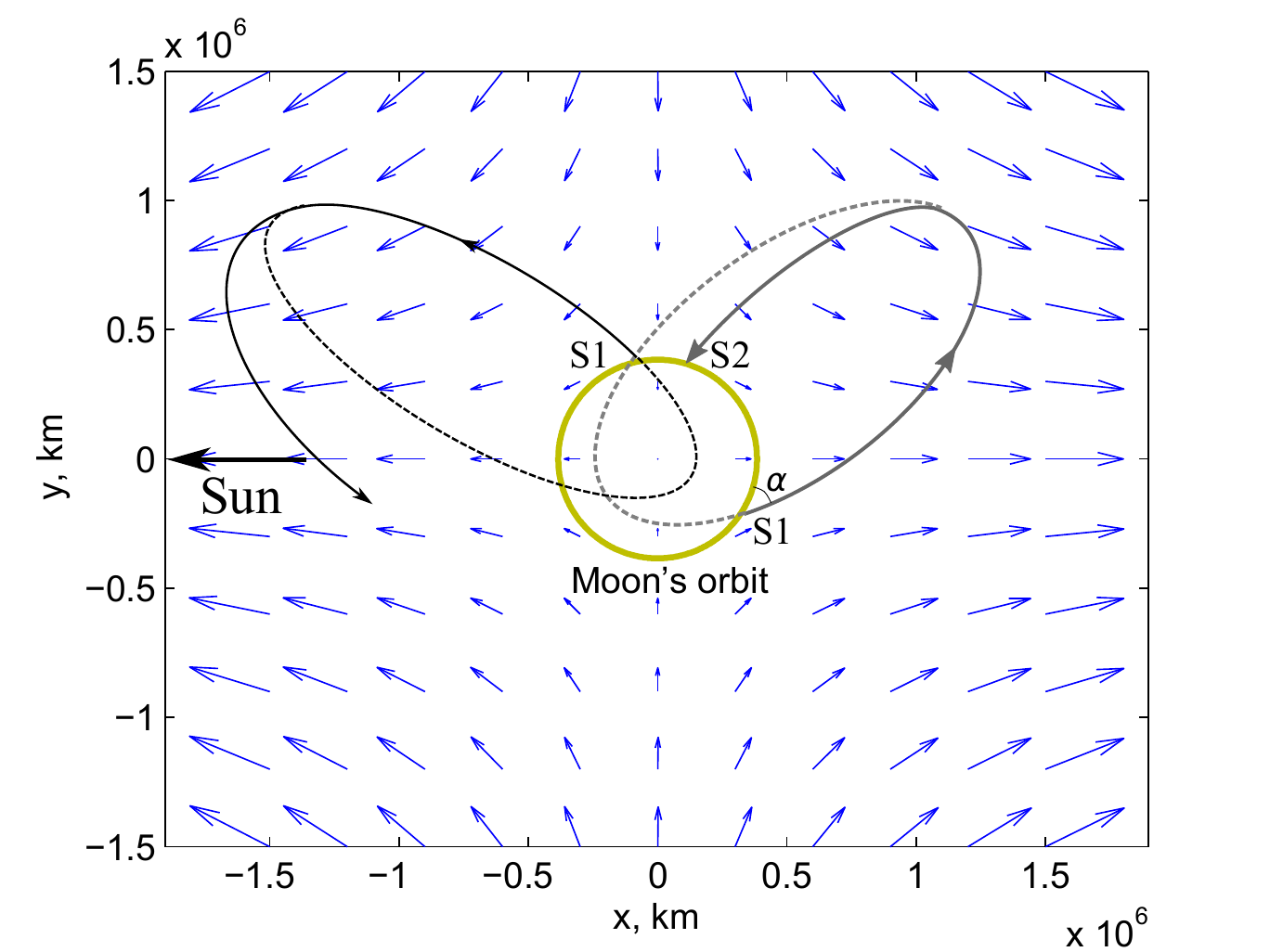}
				\caption{The solar gravitational influence on orbit states with respect to the Earth and the Moon (figure reused from Ref.~\cite{Chen2016,Chen2017}).}
				\label{fig:SolarPert}
			\end{center}
		\end{figure}

	For a trajectory to be seen as an Earth orbit, the $v_{\infty}$ with respect to the Moon, $v_{\infty\textbf{M}}$, remains constant, if solar gravity is not taken into account. In the presence of solar perturbations, $v_{\infty\textbf{M}}$ will be changed as the ``Earth" orbit reaches a far region where solar perturbations are significant. To illustrate, Fig.~\ref{fig:SolarPert} schematically depicts the solar tidal force in the Earth-centered inertial frame and its effect on orbits around the Earth.  As shown, after the first lunar swingby (S1), or an outward crossing of the Moon's orbit, the resulting orbit with its apogee far away in the 1st or 3rd quadrants will experience posigrade deceleration near the apogee, and come back to a second lunar swingby (S2) with a larger encounter angle $\alpha$ with respect to the Moon's orbit, and can even be in the retrograde direction. Even though the velocity is decreased (considering the posigrade situation), $v_{\infty\textbf{M}}$ is still increased because of the increased $\alpha$. This mechanism is further clarified via the Swingby-Jacobi graph introduced in the next section. In contrast, an orbit with its apogee in the 2nd or 4th quadrants will experience acceleration, resulting in a decreased $v_{\infty\textbf{M}}$ at S2 or missing the Moon's orbit. 
		Such Sun-perturbed Moon-to-Moon transfers (hereafter referred to as SPMT or Sun-perturbed M-M transfers) can be employed to improve the lunar encounter condition for a gravity-assisted escape, as planned for the NOZOMI and DESTINY missions~\cite{KAWAGUCHI2003189,Chen2015,Chen2016}. Furthermore, they can be used for an inexpensive insertion into a lunar orbit, as applied in the HITEN, ARTEMIS-THEMIS, EQUULEUS, and Lunar Flashlight missions~\cite{UESUGI1991347, FOLTA2012237, GarciaYarnoz2016, LunarFlashlight2020}. 
		
		\subsection{Swingby-Jacobi graph}
\begin{figure}[h!t]
			\begin{center}
				\includegraphics[width=0.8\hsize]{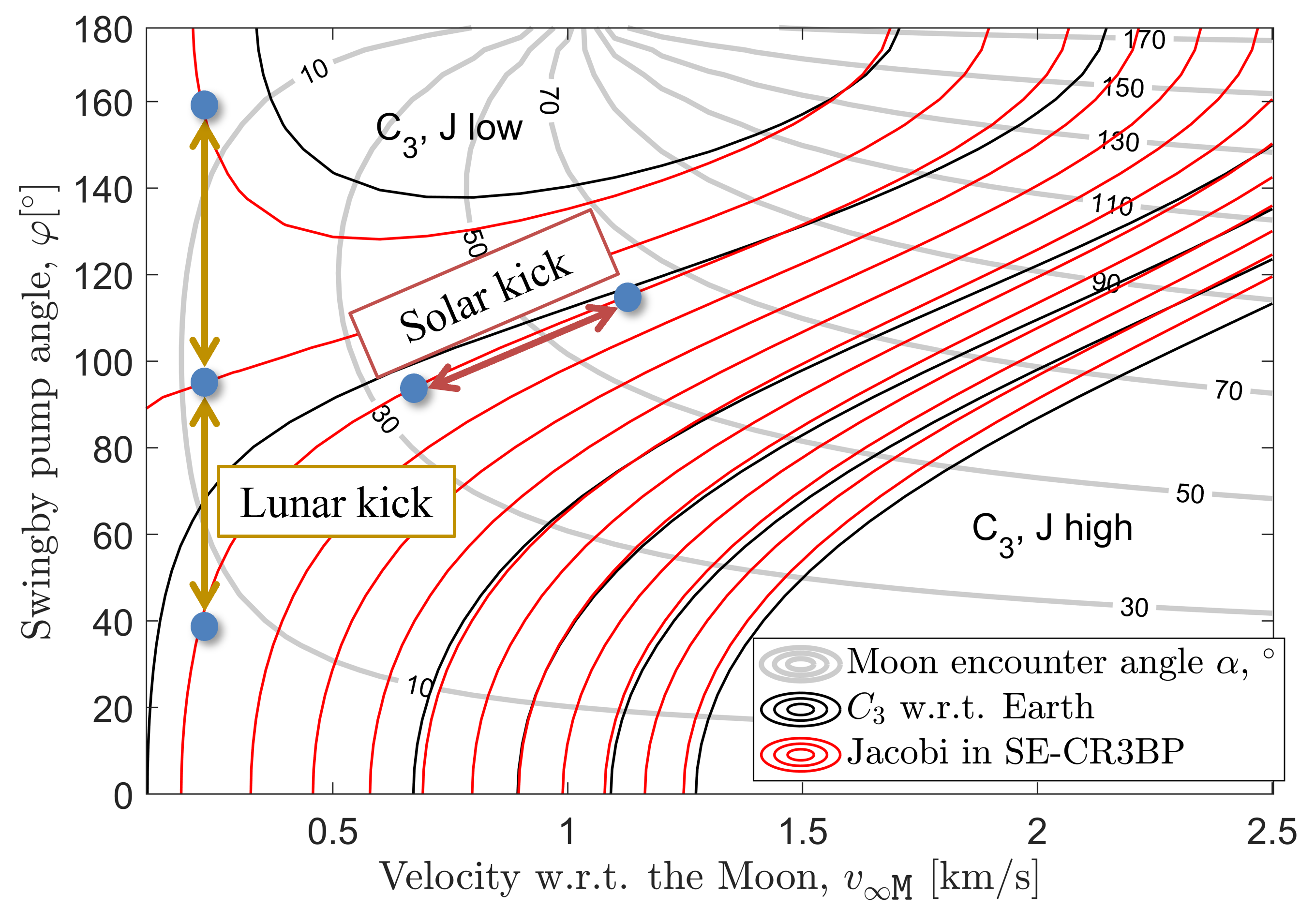}
				\caption{Swingby-Jacobi graph}
				\label{fig:jacobi}
			\end{center}
		\end{figure}

		Given \(v_{\infty\text{M}}\) and the pump angle \(\varphi\) (i.e., angle between the \({\bf{v}}_{\infty\text{M}}\) and the Moon’s velocity \({\bf{v}}_{\text{M}}\) vectors) one can compute the Jacobi integral $J$ in the SE-CR3BP. Although $J$ can vary with the position of the lunar encounter and the out-of-plane degree of \({\bf{v}}_{\infty\text{M}}\), these factors only change the Jacobi integral within an insignificant range. Figure~\ref{fig:jacobi} shows contours of $J$ on the \(v_{\infty\text{M}}\)-\(\varphi\) plane. 
		In this graph, a lunar swingby can be portrayed by the vertical motion of a state (blue dot), which changes $J$ in the Sun-Earth system. 
		The following M-M transfer can be perturbed by the Sun, while $J$ is maintained. Solar perturbations can be portrayed by motions along the Jacobi contours. 
		The Jacobi contours and the \(v_{\infty\text{M}}\)-\(\varphi\) plane constitute the basis of the ``Swingby-Jacobi'' graph.  Orbital information can be added to this graph, which enables insights into the orbit dynamics and effects on the mission. Figure~\ref{fig:jacobi} includes contours of the encounter angle $\alpha$ and osculating (i.e., at the lunar encounter) characteristic energy $C_3$ (i.e., $v^2_{\infty}$) with respect to Earth. It can be seen that in the direction that $C_3$ is slightly decreased and $\alpha$ is increased, which corresponds to the situation of posigrade deceleration in the 1st and 3rd quadrants, \(v_{\infty\text{M}}\) is increased. This visualizes and explains the mechanism by which \(v_{\infty\text{M}}\) increases while $C_3$ decreases during a Sun-perturbed transfer. To conclude, the following situations are visualized in the graph. 
		\begin{itemize}    
			\item The orbital energy with respect to Earth, represented by $C_3$, is influenced by lunisolar perturbations; 
			\item The Jacobi integral in the Sun-Earth three-body problem, represented by $J$, is influenced by lunar gravity; 
			\item The orbital energy with respect to the Moon, represented by \(v_{\infty\text{M}}\), is influenced by solar gravity;
			\item \(v_{\infty\text{M}}\) is increased after a Sun-perturbed transfer mainly in the 1st and 3rd quadrants, and decreased after a Sun-perturbed transfer in the 2nd and 4th quadrants. 
		\end{itemize}
		
		Therefore, a sequence of Sun-perturbed M-M transfers and lunar swingbys can be utilized 1) to assist escape and asteroid encounter, namely, to achieve certain $C_3$, $J$, and escape directions, 
		and 2) for asteroid capture, namely, to reduce $v_{\infty\text{M}}$ for inexpensive lunar orbit insertion. Other information can also be added to this graph. For instance, the maximum post-swingby \(C_3\) is also a function of the pre-swingby $\varphi$ and \(v_{\infty\text{M}}\), and thus can be plotted in the 	Swingby-Jacobi graph. Given an initial state, the achievable \(C_3\) after two swingbys can be revealed, which is presented in Ref.~\cite{Chen2016}. The characterization of the capacity of a Sun-driven lunar swingby sequence in Sec.~\ref{sec:jacobirange} is aided by this graph.  
		
		\subsection{Trajectory design objectives}
		It is therefore feasible to launch a spacecraft with a low initial \(v_{\infty\text{M}}\) into deep space, and to reduce the \(v_{\infty\text{M}}\) of a returned spacecraft and asteroid samples for an inexpensive insertion into a stable lunar orbit, using the Sun-driven lunar swingby sequence. 
		A low-cost initial \(v_{\infty\text{M}}\) (i.e., before acceleration via a Sun-driven lunar swingby sequence) is commonly between 200 m/s (i.e., via a low-thrust spiral orbit raising) and 800 m/s (i.e., via a direct transfer from a GTO). Therefore, for an Earth escape that targets an asteroid, the objective of trajectory design is to minimize the initial \(v_{\infty\text{M}}\) to below 800 m/s. 
		For spacecraft and asteroid capture, the objective is to achieve a final \(v_{\infty\text{M}}\) smaller than 450 m/s, enabling insertion into a mid- to high-lunar orbit at a \(\Delta v\) of around 350 m/s or a stable NRHO at 20 m/s~\cite{Oguri2020, LunarFlashlight2020,CAPSTONE2021}. Another propulsion system launched to the Earth-Moon regime years later can possibly provide this insertion $\Delta v$ while waiting to dock with the returned spacecraft and asteroid.
		
		\section{Capacity of Sun-driven lunar swingby sequences}\label{Sec:capacity}
		There are around 15,000 cataloged near-Earth asteroids. Each one has hundreds of locally optimal heliocentric transfer trajectories from and to the Earth's vicinity in any given interval of 20 years. Each Earth escape or Earth encounter condition leads to hundreds to thousands of possible lunar swingby sequences given the computed database of SPMT (introduced in Sec.~\ref{sec:database}) and a limit of two SPMT segments. Computing all possible trajectories each time the asteroid database is updated can be very time-consuming. Therefore, it is desired to reduce the search space based on the capacity of a Sun-driven lunar swingby sequence.
		\subsection{Accessible energy range}\label{sec:jacobirange}
		Looking at the Swingby-Jacobi graph, it may seem that an orbit state can be freely moved between the left side (i.e., small \(v_{\infty\text{M}}\)) and the right side (i.e., large \(v_{\infty\text{M}}\)). However, as \(v_{\infty\text{M}}\) increases, the bending angle of \(\bf{v}_{\infty\text{M}}\) becomes more restricted, and so does the Jacobi jump. When \(v_{\infty\text{M}}\) is large (i.e., $>$ 2.5 km/s), it can take many segments of SPMTs and lunar swingbys to substantially reduce or increase \(v_{\infty\text{M}}\). Additionally, when \(v_{\infty\text{M}}\) is large, $C_3$ with respect to Earth is inevitably large, and the SPMT is in a hyperbolic arc whose $ToF$ is generally longer than 200 days. Therefore, using the Sun-driven lunar swingby sequence is not efficient when $C_3$ and \(v_{\infty\text{M}}\) are large. 
		
		Time is a critical factor for space missions. In this work, the $ToF$ of an M-M transfer is limited to 200 days, and the number of patched SPMTs is limited to 2. 
		The $ToF$ constraint essentially restricts the maximum accessible $C_3$. 
		For all SPMTs limited by 200 days that have been solved for the database, $C_3$ is generally below 0 (although a few solutions reach 0.7 km$^2$/s$^2$). Therefore, a constraint is applied that limits $C_3$ to 0 before the last (or after the first) lunar swingby for escape (or capture) purposes.
		Additionally, the limit of the bending angle $\gamma$ of $\bf{v}_{\infty\text{M}}$ at a lunar swingby constrains the maximum pre-swingby or post-swingby $C_{3}$, $C_{\text{3}\max}$. $\gamma$ is limited by ~\cite{CURTIS2014405}, 
		\begin{equation}\label{eq:bendinglimit}
			{{\gamma }_{\max }}=\pi -2\times \arccos \left[ {{{G}_\text{M}}}/{({{G}_\text{M}}+{{r}_{\min }}\cdot {v}_{\infty\text{M} }^{2})}\; \right]
		\end{equation}
		where ${G}_\text{M}$ is the gravitational parameter of the Moon and ${r}_{\min }$ is the minimum flyby radius, which is set to 1838 km (i.e., 100 km above the surface of the Moon). This constraint also restricts the capacity of a Sun-driven lunar swingby sequence. 
		
\begin{figure}[h!t]
			\begin{center}
				\includegraphics[width=\hsize]{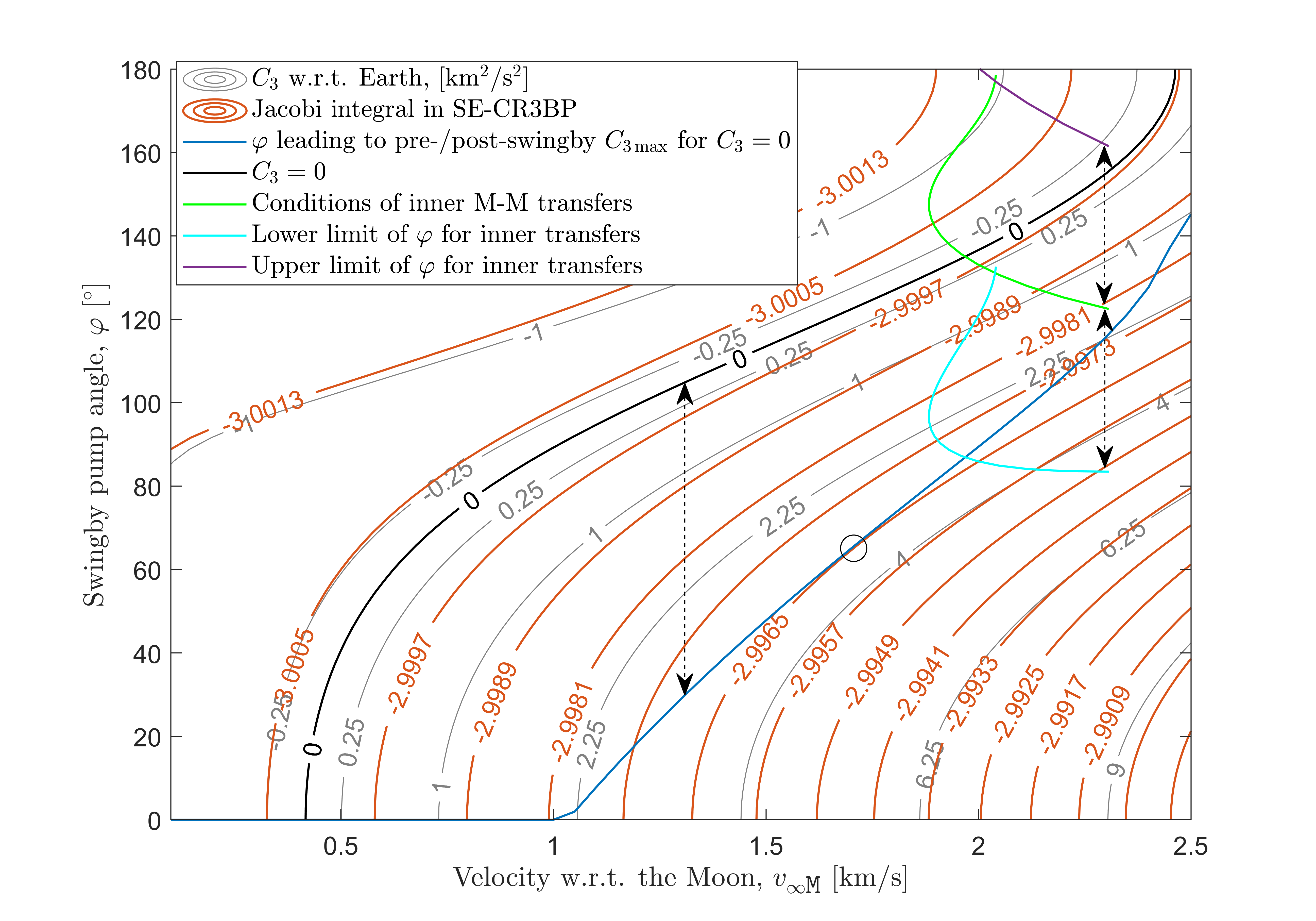}
				\caption{To reveal the capacity of a Sun-driven lunar swingby sequence in terms of the accessible Jacobi range.}
				\label{fig:JacobiRange}
			\end{center}
		\end{figure}

		To express the effects of those constraints, 
		the $C_3 = 0$ contour is highlighted on the Swingby-Jacobi graph, as shown in Fig.~\ref{fig:JacobiRange}. For $C_3 = 0$, the corresponding $\varphi$ which leads to the maximum pre-swingby or post-swingby $C_{3}$ while being constrained by $\gamma_{\max}$ is plotted as a blue line on the graph. The area bounded by these two lines indicates the accessible heliocentric orbit states after or before a Sun-driven lunar swingby sequence. The maximum accessible $J$ is obtained at the tangent point (annotated by a circle in Fig.~\ref{fig:JacobiRange}) of the $C_{3\max}$ boundary (blue) and the Jacobi contours, which is -2.9965. The loosely equivalent $C_3$ is 3.3 km$^2$/s$^2$, and the $v_{\infty}$ with respect to Earth, $v_{\infty\text{E}}$, is 1.8 km/s. In addition, orbits of low $J$ may pass by the Earth through the Sun-Earth Lagrangian $L_1$ point, where $J = -3.0009$. The accessible $J$ indicates the accessible heliocentric orbits, as the Jacobi integral can be expressed in the Tisserand form as a function of the orbital elements (for more detail, see Ref.~\cite{Kemble2006}),  which is, 
		\begin{equation}
			J=-(1-\mu )/a-2\sqrt{a(1-\mu )(1-{{e}^{2}})}\times \cos i
		\end{equation}
		where $a$, $i$, and $e$ are the semi-major axis, inclination, and eccentricity of a heliocentric orbit, respectively. 
		The reachable set $\mathcal{A}$ with a Sun-driven lunar swingby sequence can be expressed as:
		\begin{equation}
			\mathcal{A}\supset\{(a,e,i)| -3.0009<J<-2.9965\}
		\end{equation}
		
		
		The superset above means that there is another possibility to further expand the accessible heliocentric region.  
		McElrath et al. (2012) ~\cite{McElrath2012} discussed the application of the retrograde inner M-M transfer. This kind of transfer is short (i.e., 4 to 15 days) and inside the Moon's orbit, and thus is not considered as a Sun-perturbed transfer. The $v_{\infty\text{M}}$ is thus the same at the two consecutive lunar encounters. Some of these transfers are hyperbolic arcs. 
		To illustrate its useful features for the problem of interest, the solutions of inner transfers are manifested as a green curve in Fig.~\ref{fig:JacobiRange}. The lower end is associated with the minimum Earth flyby altitude, which is 220 km in this work. The orbit state can be considered to jump twice vertically as a result of the two consecutive lunar swingbys of this transfer. 
		Therefore, boundaries of the $\varphi$ leading to and resulting from the swingbys of the inner M-M transfers are also plotted in Fig.~\ref{fig:JacobiRange}. The orbit state above the lower $\varphi$ boundary (cyan) can be transferred from or to a state with $C_3 < 0$ with the two lunar swingbys at the ends of a short inner transfer. 
		Therefore, this kind of transfer provides another accessible area in the heliocentric region, which is a triangle-like area outside the aforementioned blue boundary. It can be read out that the accessible $J$ is up to -2.9946, which is loosely equivalent to a $C_3$ of 4.83 km$^2$/s$^2$ and a $v_{\infty\text{E}}$ of 2.2 km/s. However, the distribution of this level of $J$ is not isotropic in space. The reason is that the inner M-M transfer requires bending the escape (or incident) trajectory from (or onto) the ecliptic plane, and the constraint on the bending angle increases with velocity. This will be further analyzed in the next subsection. Then, the reachable set is a subset expressed as:
		\begin{equation}\label{eq: conservative_set}
			\mathcal{A}\subset\{(a,e,i)| -3.0009<J<-2.9946\}
		\end{equation}
		
		\subsection{Accessible direction range}\label{sec:directionrange}
		To use the inner M-M transfer, it should be connecting the lunar swingby sequence and the heliocentric trajectory, namely, as the last segment of the escape phase, or the first segment of the capture phase. This requires the last (or the first) swingby to bend the \(\bf{v}_{\infty\text{M}}\) from (or onto) the ecliptic plane given a targeted escape (or incident) \(\bf{v}_{\infty}\) with respect to Earth, \(\bf{v}_{\infty\text{E}}\), which may not be in the plane. 
		
		If only SPMTs are employed, it is also preferred to bend the last (or first) \(\bf{v}_{\infty\text{M}}\) from (or onto) the plane. As reasoned in Sec.~\ref{sec:3Ddiscussion}, three-dimensional SPMTs do not result in significant changes in $v_{\infty\text{M}}$ and thus are not efficient given the constraint on $ToF$ and the number of SPMTs. Because of the bending angle constraint, the escape (or capture) condition becomes more difficult to achieve as the targeted declination of \(\bf{v}_{\infty\text{E}}\) increases. It is thus necessary to derive the relationship between the accessible ranges of the magnitude \({v}_{\infty\text{E}}\) and the declination $\delta$ of \(\bf{v}_{\infty\text{E}}\).
	
\begin{figure}[h!t]
			\begin{center}
				\includegraphics[width=0.6\hsize]{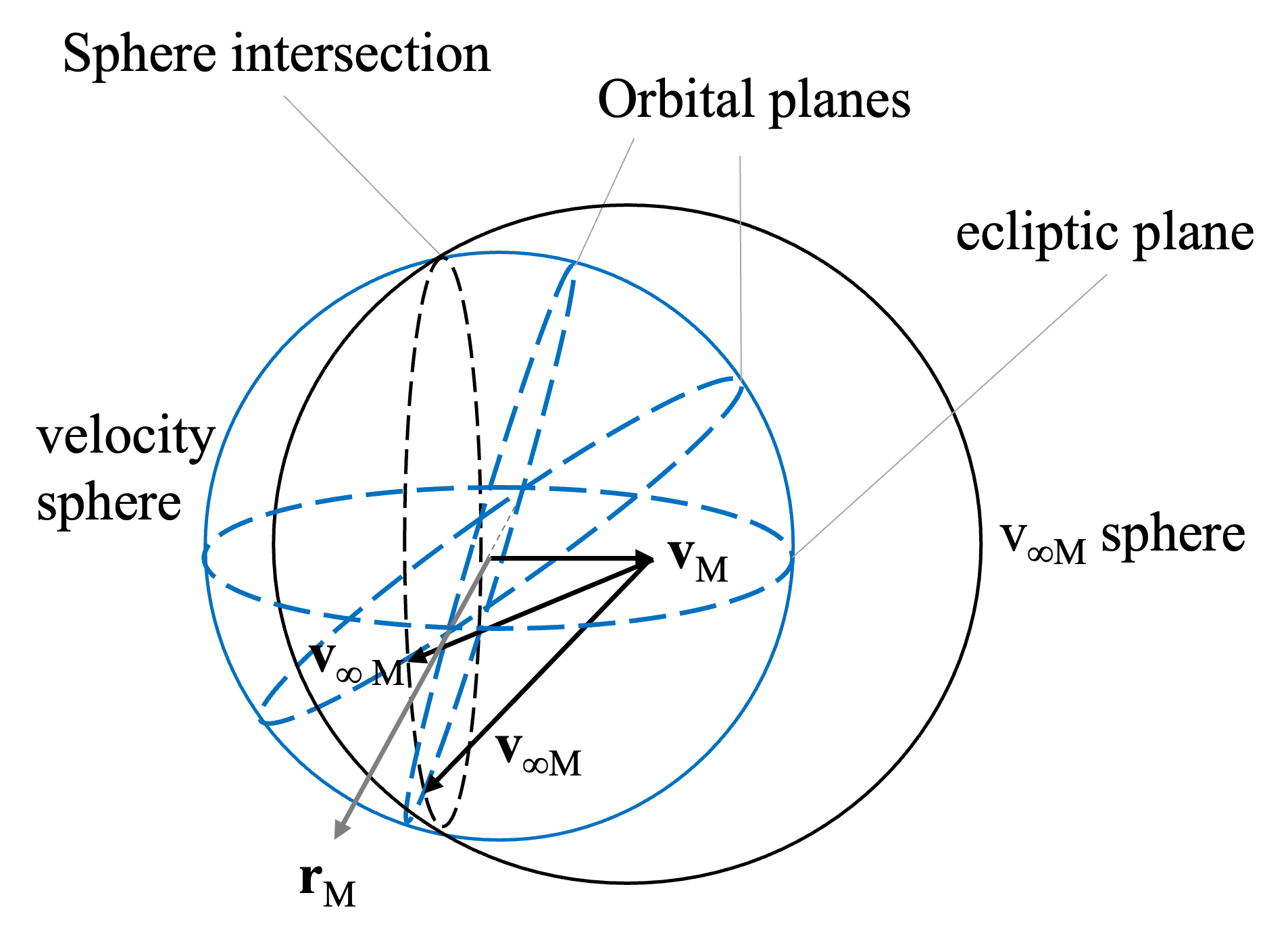}
				\caption{Lunar encounter situations for specified velocity and \(v_{\infty\text{M}}\).}
				\label{fig:declination}
			\end{center}
		\end{figure}

		The escape and incident $\bf{v}_{\infty\text{E}}$ can be related to an Earth orbit. Note that the solar perturbation is not considered in the escape and incident arc (i.e., connecting the heliocentric phase and the swingby sequence), because of its short effect and for the sake of simplicity. The corresponding Earth orbit has an inclination ranging from $\delta$ to 90$^\circ$. The inclination is determined by the difference between the ascension of $\bf{v}_{\infty\text{E}}$ and the lunar phase 
		at the lunar encounter. For the same \(v_{\infty\text{E}}\) and thus the same velocity at the lunar encounter, and the same \(v_{\infty\text{M}}\), a low-inclination orbit results in a lower declination of $\bf{v}_{\infty\text{M}}$ than a high-inclination orbit (see illustration in Fig.~\ref{fig:declination}). It is more difficult to bend the $\bf{v}_{\infty\text{M}}$ of a higher declination from (or onto) the plane.  
		
\begin{figure}[h!t]
			\begin{center}
				\includegraphics[width=0.6\hsize]{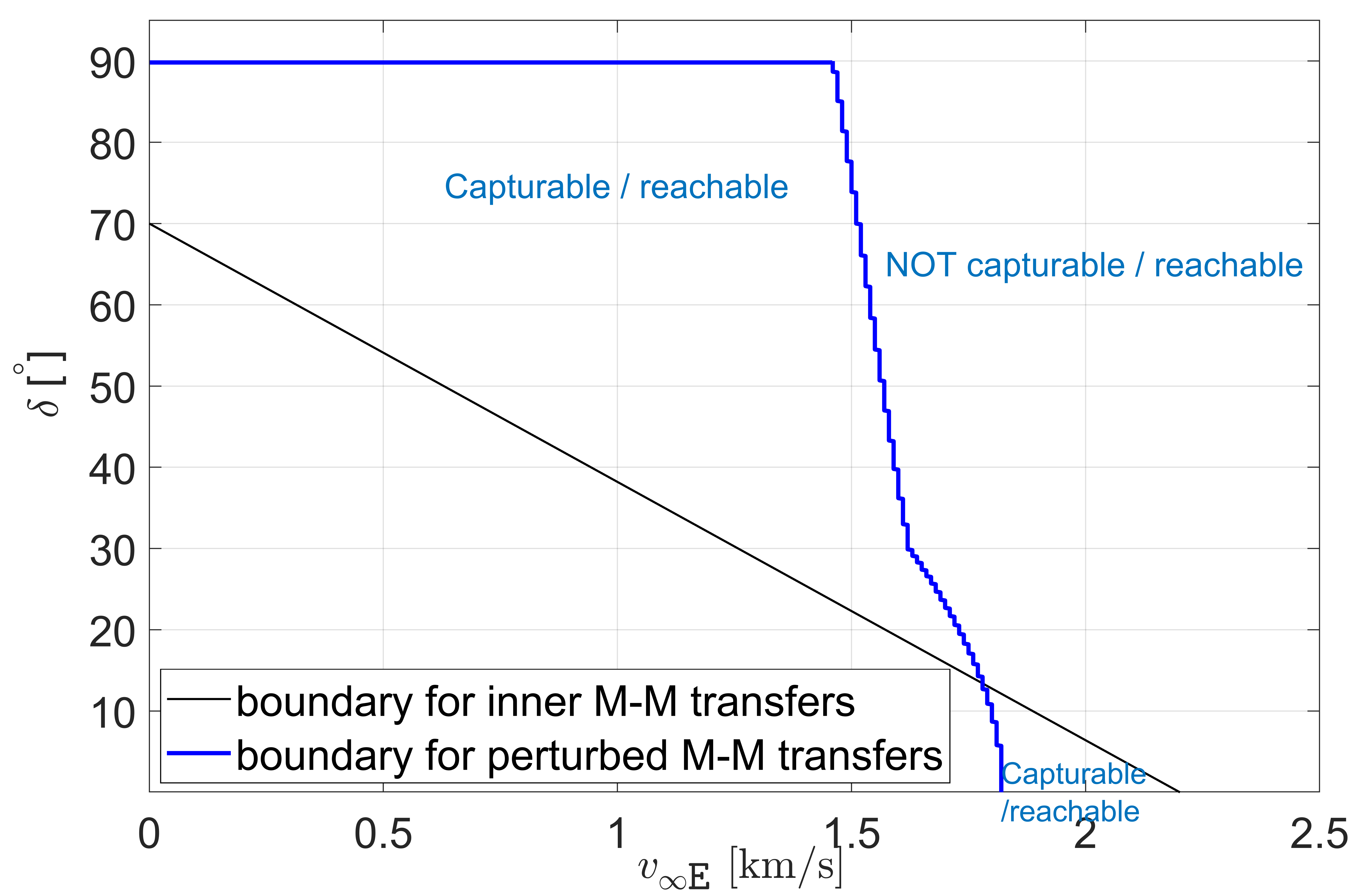}
				\caption{Accessible domain of the \(\bf{v}_{\infty\text{E}}\) magnitude and its declination.}
				\label{fig:direction}
			\end{center}
		\end{figure}

		To examine the accessible situation for an escape (incident) vector of certain $v_{\infty\text{E}}$ and $\delta$, one can vary the lunar encounter position, compute the orbit state at the encounter (i.e., see Appendix) and the bending angle limit (~Eq.~\ref{eq:bendinglimit}), and examine whether the orbit state at the encounter can be bent from (or to) the ecliptic plane with $C_3 < 0$. By examining various lunar encounter phases, $v_{\infty\text{E}}$, and $\delta$, a two-dimensional accessible area is revealed, as shown in Fig.~\ref{fig:direction}.  In addition, the situation with one inner M-M transfer in the sequence is considered. The possibility of this situation is dependent on whether the state can be bent from (or onto) the solution of the inner transfer (i.e., depicted by the green line in Fig.~\ref{fig:JacobiRange}). The corresponding boundary is also shown in Fig.~\ref{fig:direction}. Let the combined accessible domain in Fig.~\ref{fig:direction} be denoted by a set $\mathcal{B}$. The reachable set with a planar Sun-driven lunar swingby sequence, $\mathcal{A}_\text{p}$, can be expressed as,
		\begin{equation}\label{eq: exact_set}
			\mathcal{A}_\text{p} = \{(a,e,i)|(v_{\infty\text{E}}, \delta)\in\mathcal{B}\}
		\end{equation}
		While the orbital elements can be loosely mapped from $v_{\infty\text{E}}$ and $\delta$, because $\mathcal{B}$ cannot be expressed as a single function, characterizing Eq.~\ref{eq: exact_set} into explicit functions is not specifically performed in this work. Instead, Eq.~\ref{eq: conservative_set} is used to pre-select potential candidates, and Eq.~\ref{eq: exact_set} is used in a later step to verify the true accessibility, as is demonstrated in Sec.~\ref{Sec:results}.

		Figure~\ref{fig:direction} also shows that a Sun-driven lunar swingby sequence (constrained by $ToF$ and the number of SPMTs) can at least achieve or absorb a $v_{\infty\text{E}}$ of up to 1.46 km/s in all directions. The corresponding  $C_3$ is $\SI{2.13}{\kilo\meter^2/\second^2}$, and the loosely equivalent $J$ is -2.9980. Thus, $\mathcal{A}_\text{p}$ contains the subset associated with this Jacobi range, which is expressed as, 
		\begin{equation}
			\mathcal{A}_\text{p}\supset\{(a,e,i)| -3.0009<J<-2.9980\}
		\end{equation}
		
		\subsection{Potentially accessible asteroids}
		Because a spacecraft does not generally rendezvous with an asteroid, a $\Delta v$ to stop it at the asteroid, denoted by $\Delta v_{\text{stop}}$, is required. Similarly, an asteroid does not commonly reach the Earth's vicinity without a $\Delta v$ to return it, denoted by $\Delta v_{\text{retn}}$. Nevertheless, asteroids whose orbits are outside the reachable set cannot be reached or returned with a small $\Delta v$ as the expected large velocity difference at the intersection needs to be canceled. Therefore, the characterized capacity of a Sun-driven lunar swingby sequence also indicates the easily accessible targets. To show the increased capacity with the proposed method, a comparison with the methods using lunar swingbys and the Sun-Earth libration dynamics on the accessible Jacobi and the corresponding number of potential targets in the current asteroid database is presented in Table~\ref{tab:Comparision}. 
		
		\begin{table}[h!]
			\caption{Comparison of different gravity-assist methods}\label{tab:Comparision}
			\centering
			\begin{threeparttable} 
				\begin{tabular}{P{0.20\linewidth}P{0.20\linewidth}|P{0.25\linewidth}|P{0.2\linewidth}}\toprule
					\multicolumn{2}{P{0.4\linewidth}|}{Method}   & Accessible Jacobi & Number of candidates~\tnote{d} \\\midrule
					\multicolumn{2}{P{0.4\linewidth}|}{Lunar swingbys}   & {[}-3.0009, -2.9962{]}~\tnote{a}  & 491                                             \\\hline
					\multicolumn{2}{P{0.4\linewidth}|}{S-E libration-point dynamics}     & {[}-3.0009, -2.9992{]}~\tnote{b}  & 168                                            \\\hline
					\multicolumn{1}{P{0.20\linewidth}|}{\multirow{2}{\linewidth}{Sun-driven lunar swingby sequence}} & only SPMTs & {[}-3.0009, -2.9965{]}~\tnote{c} & 456 \\\cline{2-4}
					\multicolumn{1}{P{0.20\linewidth}|}{} & with an inner transfer & {[}-3.0009, -2.9946{]}~\tnote{c} & 657 \\
					\bottomrule                                          
				\end{tabular}
				\noindent
				\begin{tablenotes}
					\item[a] \footnotesize Gong and Li (2015) ~\cite{Gong2015} 
					\item[b] Sanchez and Garcia Yarnoz (2016) ~\cite{Sanchez2016}
					\item[c] The present work and Chen (2017) ~\cite{Chen2017}
					\item[d] Based on the database of the Minor Planet Center:  \href{https://minorplanetcenter.net/iau/MPCORB/MPCORB.DAT}{https://minorplanetcenter.net/iau/MPCORB/MPCORB.DAT}, accessed on 2022 Aug. 15.
				\end{tablenotes}
			\end{threeparttable}
	\end{table} 
	
	$\Delta v_{\text{stop}}$ and $\Delta v_{\text{retn}}$ are limited to a small amount, such that the effect of the gravity assists is still dominant and we do not depart from the idea of easily reachable and retrievable asteroids. 
	For reaching the target, a $\Delta v_{\text{stop}}$ budget of 1 km/s is assumed to be available. For retrieving the spacecraft and samples from the target, a $\Delta v_{\text{retn}}$ budget of 500 m/s is assumed to be available. These $\Delta v$ budgets can also enable low energy asteroids to reach Sun-Earth $L_1$ and pass by the Earth, and thus alter the lower bound of the accessible Jacobi range. After subtracting the kinetic energy attributed to the $\Delta v$ budget from the $J$ at $L_1$,  the lower bound of the $J$ of a reachable asteroid becomes -3.0020, and that of a retrievable one becomes -3.0012. 
	These Jacobi ranges are applied in the first place to exclude the definitely inaccessible asteroids.

	\section{Trajectory design}
	\subsection{Heliocentric transfer phase}\label{Sec:helio}
	As a Sun-driven lunar swingby sequence can reach or absorb a $v_{\infty\text{E}}$ of 1.46 km/s at the least (see Fig.~\ref{fig:direction}), and $\Delta v_{\text{stop}}$ and $\Delta v_{\text{retn}}$ provided by propulsion systems are limited to 1 km/s and 500 m/s, respectively, minimizing the $\Delta v$ cost in the heliocentric transfer phase is crucial. A Lambert solver for the two-body problem and the Matlab Fmincon routine are used to find low-$\Delta v$ transfer trajectories between the Earth and asteroids. The free variables, the $\Delta \bf{v}$ vector, and epochs of Earth departure, asteroid arrival, asteroid departure, and Earth arrival, are optimized. 
	The Lambert solver can consider multiple revolutions and return multiple solutions. In this work, up to three revolutions of heliocentric transfer are considered. However, the Lambert solver is not called in the optimization routine but is only used to provide initial guesses for $\Delta \bf{v}$. Starting with the initial guess, the Fmincon routine can continue to compute the optimal trajectory with the fed gradients whose expressions are available in Ref.~\cite{battin1999}, p. 467. Furthermore, using the numerical optimization routine for solving the two-point boundary problem can avoid the singularity problem in the Lambert solver when the revolution number jumps. Once an optimal transfer trajectory with $\Delta v_{\text{stop}} < 1 $ km/s (or $\Delta v_{\text{retn}}< 500$ m/s) is found, the corresponding escape (or incident) condition, $\bf{v}_{\infty\text{E}}$ as well as its declination $\delta$, is known, and the accessible domain described in Sec.~\ref{sec:directionrange} can be applied to exclude impossible transfer trajectories.
	
	\subsection{Escape and capture phases}
	The escape (or capture) phase involves a sequence of Moon-to-Moon transfers and lunar swingbys to reduce the initial (or final) \(v_{\infty\text{M}}\) to $<800$ m/s (or $450$ m/s), given the targeted escape (or incident) $\bf{v}_{\infty\text{E}}$ connected with the heliocentric transfer phase. In addition, the number of SPMT segments is preferred to be minimal, not only to limit the total flight time but also to minimize the operational risk. The number of SPMT segments is limited to 2 in this work.
	
	Given an escape (or incident) $\bf{v}_{\infty\text{E}}$, different lunar encounter positions result in different $\bf{v}_{\infty\text{M}}$ (see Appendix) and thus affect the sequence of lunar swingbys. The first (or final) lunar encounter position depends on the phase angle $\theta$ (i.e., defined by the angle of the Earth-to-Moon direction from the Sun-to-Earth direction in this work). Strictly speaking, $\theta$ depends on the departure (or arrival) date and the ephemeris of the Moon.  However, as the $ToF$ of the heliocentric phase (i.e., 250 to 1500 days) is much longer than the synodic period of the Moon (i.e., 30 days), it is assumed that a small $\Delta v$ can vary the departure (or arrival) date by 15 days without significantly affecting $\bf{v}_{\infty\text{E}}$. Therefore, $\theta$ is considered to be a variable in the exploration of lunar swingby sequences. This will allow for more options for lunar swingby sequences as well as a higher chance of meeting the $v_{\infty\text{M}}$ objective. To facilitate patching SPMT segments and lunar swingbys, a database of SPMT solutions is used. The following subsection presents the steps for generating this database.
	
	\subsubsection{Database of Sun-perturbed Moon-to-Moon Transfers}\label{sec:database}
	With significant solar perturbations, the Sun-perturbed Moon-to-Moon Transfer (SPMT) cannot be solved analytically. Nevertheless, the condition of SPMT can be defined by only two variables: $\theta$ and \(v_{\infty\text{M}}\), which are practically bounded. \(v_{\infty\text{M}}\) is bounded as the $ToF$ of an M-M transfer is limited to an acceptable time. Therefore, the grid search method is applicable.  
	The solution of SMPT includes two variables: the post-swingby direction angle $\psi$ of the post-swingby \(\bf{v}_{\infty\text{M}}^+\), and the $ToF$ for a lunar re-encounter. Lantoine and McElrath (2014)~\cite{Lantoine2014} classified the SMPT into families according to the $ToF$ range and whether the beginning and the end of the transfer trajectory go in (i.e., noted by ``i'') or out (i.e., noted by ``o'') of the Moon's orbit. M-M transfers in the same family exhibit a continuation on $\theta$, \(v_{\infty\text{M}}\), $\psi$, and $ToF$. Therefore, solutions of a family can be recovered if one solution of the family has been obtained.
	Furthermore, Garcia Yárnoz et al. (2016)~\cite{GarciaYarnoz2016} used the analytical expression of the gradient of the solution space to generate initial guesses for the neighboring grid nodes. 
	However, they also noted that as $ToF$ increases, the classification of families becomes complicated. Moreover, singularity happens as the perigee of the solution reaches a small value. Thus, it is tricky to determine the profile of families in the solution space. Without full knowledge, the continuation method is likely to miss solutions.
	
\begin{figure}[h!t]
		\begin{center}
			\includegraphics[width=0.5\hsize]{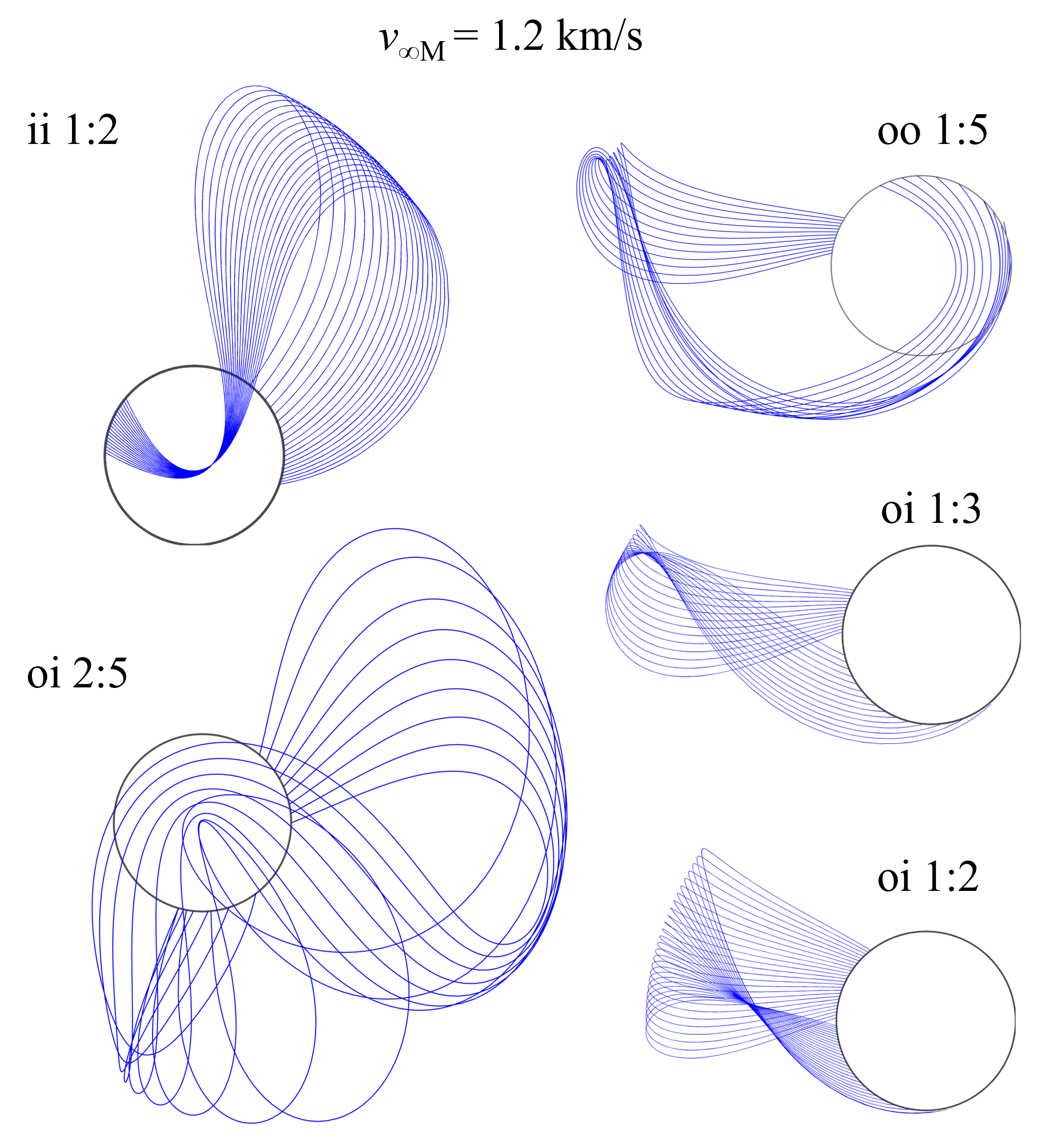}
			\caption{Examples of Sun-perturbed Moon-to-Moon transfers (synodic frame).}
			\label{fig:ExTransfers}
		\end{center}
	\end{figure}

	Considering that the database of SPMT, once established, can be accessed repeatedly, efforts and time to establish such a database are not the primary concern. Therefore, despite its tediousness, the following effort has been made to ensure completeness of the database. Solutions for every grid node defined by $\theta$ and \(v_{\infty\text{M}}\) are computed regardless of the classification of the family. However, there are infinite solutions if $ToF$ is not limited. In this work, $ToF$ is limited to 200 days. In addition, when \(v_{\infty\text{M}}\) is smaller than 0.4 km/s, the gravity loss becomes significant and the model of patching M-M transfers in the SE-CR3BP cannot hold very true. When \(v_{\infty\text{M}}\) is greater than 2.2 km/s, the trajectory is mostly hyperbolic and cannot come back to the Moon within 200 days. Therefore, a mesh grid is generated by sampling \(v_{\infty\text{M}}\) from 0.4 to 2.2 km/s spaced in intervals of 0.1 km/s and $\theta$ from 0° to 360° spaced in intervals of 2°. To solve for the SPMT for a given grid node, a combination of adaptable-step searches on $\psi$ and  differential correction on $\psi$ and $ToF$ is used (see Ref.\cite{Chen2017} for more detail). The constraint that the perigee must be greater than 6600 km for the transfer trajectories is applied. In addition, multiple-revolution (multi-rev) transfers can also be computed with this method. Figure~\ref{fig:ExTransfers} presents example SPMT with an initial \(v_{\infty\text{M}} = 1.2~\si{\kilo\meter/\second}\). They are grouped into the families defined by Lantoine and McElrath (2014)~\cite{Lantoine2014}, where the resonance $p:q$ indicates that the transfer trajectory makes $p$ revolutions (i.e., defined by the number of apogees) while the Moon rotates $q$ revolutions. Changes in lunar encounter situations at the ends of the Sun-perturbed transfers can be clearly seen. 
	

The solution space where the SPMT encounters the Moon tangentially and where $ToF$ is long is highly nonlinear, and therefore the computation is very sensitive to the initial guess. There are solutions missed in the previous step. Thus, in the second step, a scan is performed. To be specific, the continuation method is used to generate solutions for neighboring grid nodes based on every existent solution. When a new solution is found, it is added to the database. With all these efforts, the computed database is believed to be almost complete. 

\section{Results}\label{Sec:results}
\begin{figure}[h!t]
	\begin{center}
		\includegraphics[width=0.5\hsize]{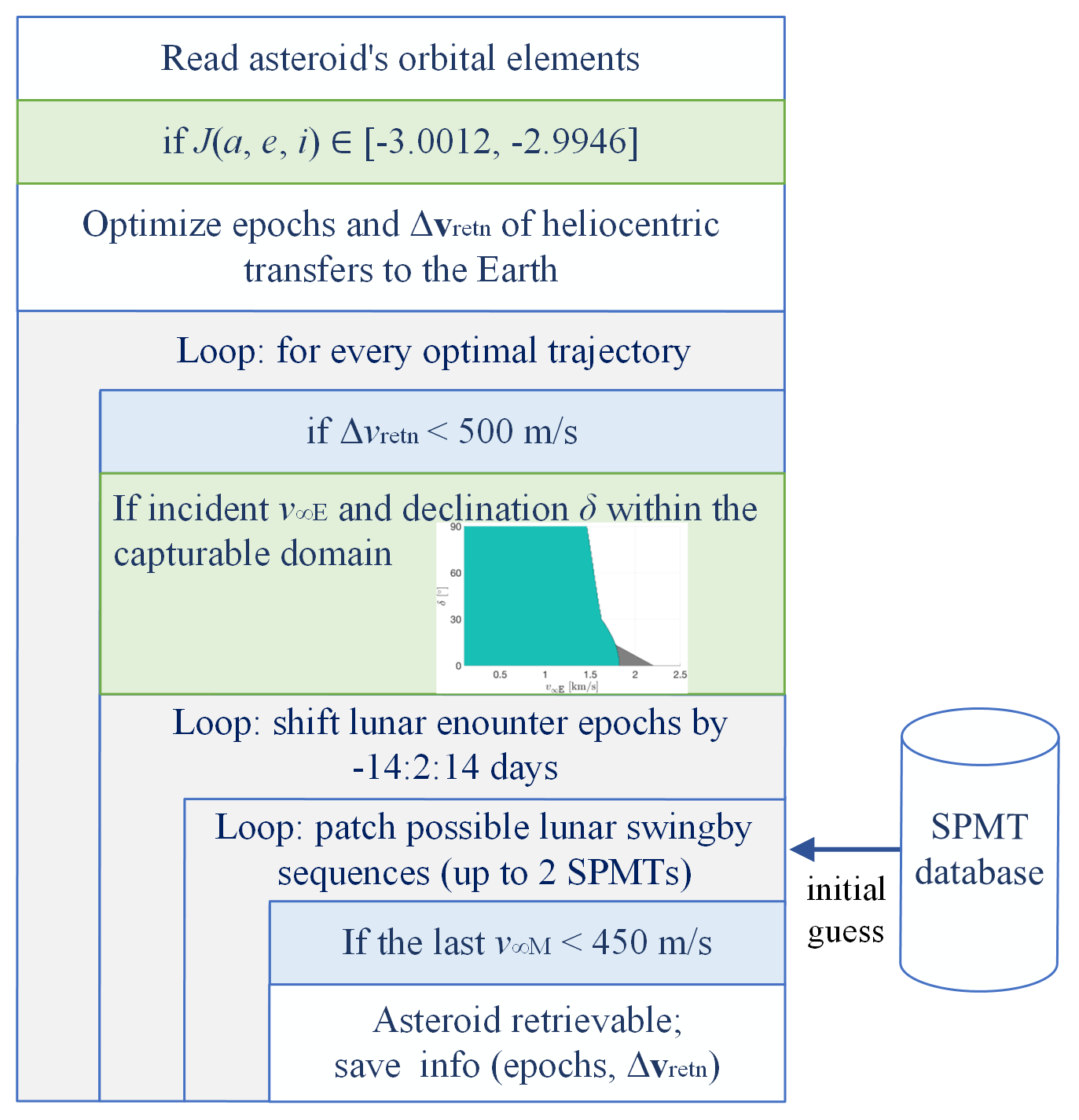}
		\caption{Workflow for finding retrievable asteroids and feasible trajectory options considering the capacity of a Sun-driven lunar swingby sequence (green blocks).}
		\label{fig:workflow}
	\end{center}
\end{figure}

Figure~\ref{fig:workflow} summarizes the workflow for finding retrievable asteroids and possible mission and trajectory profiles, where filters defined by the reachable sets, namely, the accessible Jacobi range and $\bf{v}_{\infty\text{E}}$ domain, are applied. The workflow for asteroid rendezvous is similar to that in Fig.~\ref{fig:workflow} and thus is not presented. Some of the candidates enclosed by the accessible Jacobi range may be found inaccessible for a given interval (e.g., 2023-2043 used in this work) because of unachievable escape (or incident) conditions, or no sequence of lunar swingbys found to meet the initial (or final) $v_{\infty\text{M}}$ requirements. Nevertheless, a number of accessible asteroids and feasible mission options are identified. 

Table~\ref{tab:ReachableAsteroids} summarizes the 48 asteroids that spacecraft can rendezvous with before 2043. Each asteroid can be reached with a Sun-driven lunar swingby sequence and $\Delta v_{\text{stop}}<1$ km/s. The orbital elements, the number of observations $n_{\text{obs}}$, number of oppositions, $n_{\text{opp}}$, absolute magnitude, $H$, suspected diameter, $D$, and orbit class of the asteroids are also listed for reference of mission planners. Note that for each listed asteroid, there are one to hundreds of heliocentric transfer options whose escape conditions can be met by at least one option of the Sun-driven lunar swingby sequence. Not all feasible mission profiles can be presented in this paper, but options prioritizing the flight time $TOF$ and the asteroid observation time $T_{\text{obs}}$ are selected and presented in Table~\ref{tab:FlybySummary}. $T_{\text{obs}}$ represents the minimum period of observing the asteroid in case the required $\Delta v_{\text{stop}}$ cannot be fully paid. It is related to $H$, the velocity vector of the spacecraft with respect to the asteroid, and the camera detection magnitude~\cite{2007JBAA..117..342D}. Herein, a camera magnitude of 12 is adopted, which was used by the low-cost micro-probe PROCYON~\cite{Yam2014}. As the table indicates, even without $\Delta v_{\text{stop}}$, the Sun-driven lunar swingby sequence can still permit asteroid flybys for closely observing asteroids over hours to more than one week (e.g., for 1991 VG, 2000 SG344, 2014 YD, 2017 TB18, and 2017 BN93). 

\begin{longtable}[htp]{l*{7}{r}l}
	\caption{Easily reachable asteroids before 2043}\label{tab:ReachableAsteroids}\\
	\toprule
	Asteroid Des. ID & $H$     & $i$ [$^\circ$] & $e$     & $a$ [AU] &  $D$ [m] & $n_{\text{obs}}$ & $n_{\text{opp}}$ & Class       \\\midrule
	1991 VG    & 28.3 & 1.4 & 0.052 & 1.032 & 10       & 66   & 3 & Apollo      \\
	2000 LG6   & 29.0 & 2.8 & 0.111 & 0.917 & 7        & 13   & 1 & Aten        \\
	2000 SG344 & 24.7 & 0.1 & 0.067 & 0.977 & 37       & 31   & 2 & Aten        \\
	2001 GP2   & 26.4 & 1.3 & 0.072 & 1.035 & 25       & 58   & 2 & Apollo      \\
	2005 QP87  & 27.7 & 0.3 & 0.175 & 1.233 & 10-15    & 85   & 1 & Amor        \\
	2006 RH120 & 29.5 & 0.6 & 0.024 & 1.033 & 6        & 133  & 2 & Apollo      \\
	2006 UB17 & 26.3 & 2.0  & 0.103 & 1.140 & 15-33 & 28 & 1  & Amor \\    
	2007 UN12  & 28.7 & 0.2 & 0.059 & 1.049 & 9        & 132  & 2 & Apollo      \\
	2007 VU6   & 26.5 & 1.2 & 0.091 & 0.976 & 13-29    & 38   & 1 & Aten        \\
	2008 EA9   & 27.7 & 0.4 & 0.075 & 1.050 & 14       & 56   & 1 & Apollo      \\
	2008 HU4   & 28.3 & 1.4 & 0.056 & 1.071 & 7        & 77   & 2 & Amor        \\
	2008 UA202 & 29.4 & 0.3 & 0.068 & 1.033 & 5        & 16   & 1 & Apollo      \\
	2009 BD    & 28.1 & 1.3 & 0.051 & 1.062 & 10       & 178  & 3 & Apollo      \\
	2009 YR    & 28.0 & 0.7 & 0.110 & 0.942 & 12       & 29   & 1 & Aten        \\
	2010 JW34  & 28.1 & 2.3 & 0.055 & 0.981 & 12       & 55   & 1 & Aten        \\
	2010 UJ    & 26.2 & 0.4 & 0.092 & 0.950 & 20       & 12   & 1 & Aten        \\
	2010 UE51  & 28.3 & 0.6 & 0.060 & 1.055 & 7        & 175  & 1 & Apollo      \\
	2010 VQ98  & 28.2 & 1.5 & 0.027 & 1.023 & 11       & 49   & 1 & Apollo      \\
	2011 BL45  & 27.1 & 3.1 & 0.021 & 1.038 & 18       & 24   & 1 & Apollo      \\
	2011 BQ50  & 28.0 & 0.4 & 0.098 & 0.950 & 7-15     & 25   & 1 & Amor        \\
	2011 MD    & 28.0 & 2.5 & 0.037 & 1.056 & 5.19     & 1487 & 1 & Amor/Apollo \\
	2011 UD21  & 28.5 & 1.1 & 0.030 & 0.979 & 10       & 83   & 1 & Aten        \\
	2012 TF79  & 27.4 & 1.0& 0.038 & 1.050 & 16       & 63   & 1 & Apollo      \\
	2012 WR10  & 28.7 & 0.3 & 0.112 & 1.085 & 9        & 42   & 1 & Apollo      \\
	2013 BS45  & 25.9 & 0.8 & 0.084 & 0.992 & 31       & 92   & 2 & Aten        \\
	2013 EC20  & 29.0 & 1.3 & 0.121 & 1.113 & 7        & 55   & 1 & Apollo      \\
	2013 RZ53  & 31.1 & 2.1 & 0.028 & 1.017 & 2        & 31   & 1 & Apollo      \\
	2014 DJ80  & 26.3 & 3.0 & 0.067 & 0.977 & 27       & 30   & 1 & Aten        \\
	2014 HN2   & 26.3 & 1.2 & 0.118 & 0.927 & 26       & 65   & 1 & Aten        \\
	2014 QN266 & 26.3 & 0.5 & 0.092 & 1.053 & 27       & 82   & 1 & Apollo      \\
	2014 UV210 & 26.9 & 0.6 & 0.132 & 1.155 & 20       & 62   & 2 & Apollo      \\
	2014 WU200 & 29.1 & 1.3 & 0.071 & 1.028 & 7        & 46   & 1 & Apollo      \\
	2014 WX202 & 29.6 & 0.4 & 0.059 & 1.036 & 6        & 41   & 1 & Apollo      \\
	2014 WA366 & 26.9 & 1.6 & 0.071 & 1.034 & 20       & 54   & 1 & Apollo      \\
	2014 YD    & 24.3 & 1.7 & 0.087 & 1.072 & 20       & 104  & 1 & Apollo      \\
	2015 JD3   & 25.5 & 2.7 & 0.008 & 1.058 & 38       & 37   & 1 & Amor        \\
	2015 PS228 & 28.8 & 0.4 & 0.084 & 1.057 & 9        & 38   & 1 & Apollo      \\
	2015 VC2   & 27.4 & 0.9 & 0.074 & 1.053 & 16       & 111  & 2 & Apollo      \\
	2015 XZ378 & 27.2 & 2.7 & 0.035 & 1.015 & 17       & 35   & 1 & Apollo      \\
	2015 YO10  & 26.6 & 2.4 & 0.100 & 1.122 & 23       & 28   & 1 & Apollo      \\
	2016 CF137 & 25.6 & 2.5 & 0.100 & 1.091 & 36       & 50   & 1 & Apollo      \\
	2016 RD34  & 27.7 & 2.0 & 0.035 & 1.046 & 14       & 90   & 1 & Apollo      \\
	2016 TB18  & 24.8 & 1.5 & 0.084 & 1.077 & 52       & 98   & 1 & Apollo      \\
	2016 TB57  & 26.1 & 0.3 & 0.123 & 1.102 & 29       & 137  & 1 & Apollo      \\
	2017 BN93  & 25.4 & 2.1 & 0.052 & 1.044 & 40       & 15   & 1 & Apollo      \\
	2017 FJ3   & 29.9 & 1.0 & 0.118 & 1.133 & 5        & 15   & 1 & Apollo      \\
	2017 FT102 & 29.5 & 1.5 & 0.059 & 1.038 & 6        & 79   & 1 & Apollo      \\
	2017 HU49  & 26.5 & 2.6 & 0.055 & 0.971 & 24       & 147  & 1 & Aten             \\\bottomrule
\end{longtable}

\begin{longtable}[h]{lllrrr}
	\caption{Selected asteroid flyby and rendezvous missions before 2043 prioritizing flight time and observation time}\label{tab:FlybySummary}\\\toprule
	Asteroid Des. ID & Escape from Earth & Arrival at asteroid & $ToF$ [day] & $T_{\text{obs}}$ [day] & $\Delta v_{\text{stop}}$ [m/s] \\ \midrule
	1991 VG    & 2038 NOV 29 & 2041 FEB 23 & 817  & 7.9  & 120.1 \\ \hline
	2000 LG6   & 2028 MAY 16 & 2030 MAY 10 & 724  & 0.2  & 936.9 \\ \hline
	2000 SG344 & 2028 MAY 09 & 2029 JUL 02 & 418  & 54.8 & 44.0  \\ 
	& 2028 MAY 10 & 2030 JUN 20 & 771  & 59.2 & 39.8  \\ \hline
	2001 GP2   & 2038 MAR 25 & 2039 JUL 14 & 476  & 1.5  & 694.2 \\ \hline
	2005 QP87  & 2034 AUG 24 & 2037 JAN 19 & 879 & 0.7  & 770.9 \\ \hline
	2006 RH120 & 2028 NOV 14 & 2029 OCT 01 & 320  & 1.5  & 662.7 \\ \hline
	2006 UB17  & 2033 OCT 24 & 2035 OCT 14 & 719 & 1.1  & 891.0\\ \hline
	2007 UN12  & 2035 OCT 15 & 2037 NOV 04 & 752  & 2.9  & 138.1 \\ \hline
	2007 VU6   & 2037 DEC 15 & 2039 MAR 01 & 441  & 1.0  & 920.2 \\ \hline
	2008 EA9   & 2033 NOV 25 & 2035 MAR 15 & 475  & 2.9  & 217.8 \\ \hline
	2008 HU4   & 2026 APR 25 & 2027 JUL 29 & 460  & 2.3  & 186.8 \\ 	
	& 2026 APR 25 & 2028 OCT 04 & 893  & 3.5  & 94.1  \\ \hline
	2008 UA202 & 2028 OCT 20 & 2030 SEP 22 & 702  & 1.3  & 209.2 \\ \hline
	2009 BD    & 2034 JUN 03 & 2035 MAY 07 & 338  & 1.8  & 258.3 \\ \hline
	2009 YR    & 2031 NOV 15 & 2033 APR 05 & 507 & 1.5  & 535.4 \\ \hline
	2010 JW34  & 2039 NOV 05 & 2042 JUL 27 & 995  & 1.1  & 887.1 \\ \hline
	2010 UJ    & 2034 SEP 19 & 2035 MAY 25 & 248  & 3.5  & 369.9 \\ \hline
	2010 UE51  & 2036 NOV 25 & 2037 OCT 11 & 320  & 2.6  & 390.0 \\ \hline
	2010 VQ98  & 2039 NOV 13 & 2041 OCT 01 & 688  & 4.3  & 198.7 \\ 
	& 2039 NOV 08 & 2042 SEP 05 & 1032 & 5.1  & 87.2 \\ \hline
	2011 BL45  & 2028 AUG 15 & 2029 JUL 01 & 320  & 0.8  & 751.3 \\ \hline
	2011 BQ50  & 2037 FEB 02 & 2038 MAR 09 & 400  & 0.8  & 663.0 \\ \hline
	2011 MD    & 2036 JUN 23 & 2038 SEP 10 & 809  & 3.7  & 134.4 \\ \hline
	2011 UD21  & 2039 APR 14 & 2041 JAN 14 & 641  & 0.7  & 589.8 \\ \hline
	2012 TF79  & 2026 APR 09 & 2027 DEC 01 & 601  & 1.7  & 375.5 \\ 
	& 2027 FEB 14 & 2028 DEC 20 & 675  & 1.8  & 862.8 \\ \hline
	2012 WR10  & 2038 OCT 13 & 2039 MAR 18 & 156  & 1.4  & 667.8 \\ \hline
	2013 BS45  & 2024 NOV 15 & 2026 MAY 16 & 547  & 1.5 & 977.2\\ \hline
	2013 EC20  & 2033 MAY 15 & 2035 JUN 10 & 756  & 0.1  & 983.0 \\ \hline
	2013 RZ53  & 2023 SEP 06 & 2027 NOV 28 & 1544 & 0.3  & 484.7 \\ \hline
	2014 DJ80  & 2038 AUG 14 & 2040 MAY 04 & 629  & 0.9  & 993.0 \\ \hline
	2014 HN2   & 2029 MAR 06 & 2032 MAR 11 & 1101 & 1.3  & 772.6 \\ \hline
	2014 QN266 & 2026 FEB 14 & 2027 MAY 05 & 445  & 1.4  & 820.8 \\ \hline
	2014 UV210 & 2034 NOV 14 & 2036 MAR 14 & 486  & 1.1  & 783.1 \\ \hline
	2014 WU200 & 2039 DEC 17 & 2040 SEP 17 & 275  & 2.8  & 104.7 \\ \hline
	2014 WX202 & 2035 FEB 13 & 2036 JAN 04 & 325  & 1.3  & 406.0 \\ 
	& 2034 NOV 28 & 2037 MAR 15 & 838  & 2.4  & 100.6 \\ \hline
	2014 WA366 & 2033 NOV 28 & 2034 SEP 25 & 301  & 1.3  & 650.8 \\ \hline
	2014 YD    & 2035 APR 30 & 2036 FEB 25 & 301  & 3.1  & 967.8 \\ 
	& 2024 JAN 18 & 2027 MAY 25 & 1223 & 8.2  & 348.3 \\ \hline
	2015 JD3   & 2039 MAY 04 & 2040 DEC 24 & 600  & 5.3  & 663.1 \\ \hline
	2015 PS228 & 2027 AUG 16 & 2028 JUN 06 & 295  & 0.7  & 475.0 \\  \hline
	2015 VC2   & 2029 MAR 19 & 2031 JAN 01 & 653  & 3.8  & 188.9 \\ \hline
	2015 XZ378 & 2023 MAY 17 & 2026 OCT 26 & 1258 & 0.7  & 865.3 \\ \hline
	2015 YO10  & 2028 JAN 18 & 2028 NOV 18 & 306  & 0.9  & 957.5 \\ \hline
	2016 CF137 & 2040 JAN 31 & 2040 DEC 02 & 306  & 1.9  & 776.0 \\ \hline
	2016 RD34  & 2031 SEP 12 & 2033 AUG 11 & 699  & 6.2  & 87.4  \\ \hline
	2016 TB18  & 2026 MAR 29 & 2027 FEB 01 & 309  & 6.8  & 309.4 \\ \hline
	2016 TB57  & 2024 JAN 20 & 2025 JAN 21 & 367  & 2.3  & 542.5 \\ \hline
	2017 BN93  & 2032 AUG 14 & 2033 NOV 21 & 464  & 7.2  & 725.4 \\ \hline
	2017 FJ3   & 2034 MAR 04 & 2035 JUN 02 & 455  & 0.3  & 831.3 \\ \hline
	2017 FT102 & 2034 SEP 29 & 2035 JUN 13 & 257  & 0.3  & 863.0 \\ \hline
	2017 HU49  & 2039 MAY 15 & 2040 MAR 05 & 295  & 3.6  & 669.4 \\ \bottomrule
\end{longtable}

Regarding asteroid sample return and retrieval, 25 targets are found to be possible by 2043. Each can be retrieved with a Sun-driven lunar swingby sequence and a $\Delta v_{\text{retn}} < $ 500 m/s. Again, not all feasible mission profiles can be presented, but the representative missions prioritizing $\Delta v_{\text{retn}}$ are presented in Table \ref{tab:RetrievalSummary}. Readers interested in all mission options may contact the author. All retrievable targets are included in the list of easily reachable asteroids, and thus readers can refer to Table~\ref{tab:ReachableAsteroids} for general information on these targets.
Among the listed missions, $\Delta v_{\text{retn}}$ is particularly small for the 1991 VG mission, which is 32 m/s. Even though 1991 VG is 5 to 10 times heavier than 2006 RH120 and 2008 UA202, the required total impulse to return 1991 VG to the vicinity of the Earth is still the lowest of all listed missions. 
The mass of 1991 VG, which measures 10 m across, is estimated to be 1500 tons (i.e., at an assumed density of 2.8 g/cm$^3$). The required total impulse is $\SI{4.6e4}{\kilo\newton\cdot\second}$, which is within the capacity of state-of-the-art heavy-lift boosters ~\cite{Landau2013}.

\begin{table}[h!]
	\centering
	\caption{Selected asteroid sample-return and retrieval missions prioritizing $\Delta v_{\text{retn}}$}\label{tab:RetrievalOnly}
	\begin{tabular}{lllr}\toprule
		Asteroid   & Departure for Earth & Arrival at Earth & $\Delta v_{\text{retn}}$ [m/s] \\\midrule
		1991 VG    & 2036 JAN 07 & 2038 DEC 03 & 31.6  \\
		2000 SG344 & 2028 MAR 13 & 2030 OCT 07 & 182.9 \\
		2006 RH120 & 2028 APR 13 & 2028 NOV 12 & 143.6 \\
		2007 UN12  & 2031 NOV 09 & 2035 OCT 26 & 75.8  \\
		2008 EA9   & 2030 FEB 11 & 2034 FEB 08 & 70.0  \\
		2008 HU4   & 2033 FEB 22 & 2037 APR 25 & 148.7 \\
		2008 UA202 & 2025 DEC 02 & 2028 OCT 05 & 161.8 \\
		2009 BD    & 2031 APR 22 & 2035 JUN 04 & 142.6 \\
		2010 JW34  & 2038 JUN 03 & 2041 MAY 10 & 396.5 \\
		2010 UJ    & 2032 MAY 09 & 2033 DEC 06 & 389.6 \\
		2010 UE51  & 2035 JUN 08 & 2036 NOV 06 & 131.9 \\
		2010 VQ98  & 2039 FEB 06 & 2040 NOV 06 & 65.4  \\
		2011 UD21  & 2038 JUL 25 & 2041 OCT 11 & 68.0  \\
		2012 TF79  & 2026 MAR 03 & 2027 OCT 14 & 346.4 \\
		2014 WU200 & 2035 JUN 22 & 2038 DEC 18 & 104.0 \\
		2014 WX202 & 2030 JUL 26 & 2033 NOV 24 & 94.5  \\
		2014 WA366  & 	 2030 APR 23 & 2034 JUL 07 & 489.7\\
		2014 YD    & 2034 JAN 21 & 2036 JAN 19 & 353.2 \\
		2015 PS228 & 2026 APR 23 & 2029 MAR 11 & 489.7 \\
		2015 VC2   & 2025 JAN 05 & 2029 FEB 19 & 281.2 \\
		2016 RD34  & 2029 SEP 09 & 2032 SEP 12 & 151.6 \\
		2016 TB18  & 2032 JUL 27 & 2036 MAR 19 & 234.4 \\
		2017 BN93  & 2027 AUG 24 & 2031 AUG 08 & 166.7 \\
		2017 FJ3   & 2028 MAR 16 & 2029 MAR 13 & 479.9 \\
		2017 FT102 & 2030 SEP 09 & 2034 APR 19 & 485.4\\ \bottomrule
	\end{tabular}
\end{table}

\begin{figure}[h!t]
	\centering\includegraphics[width=\textwidth]{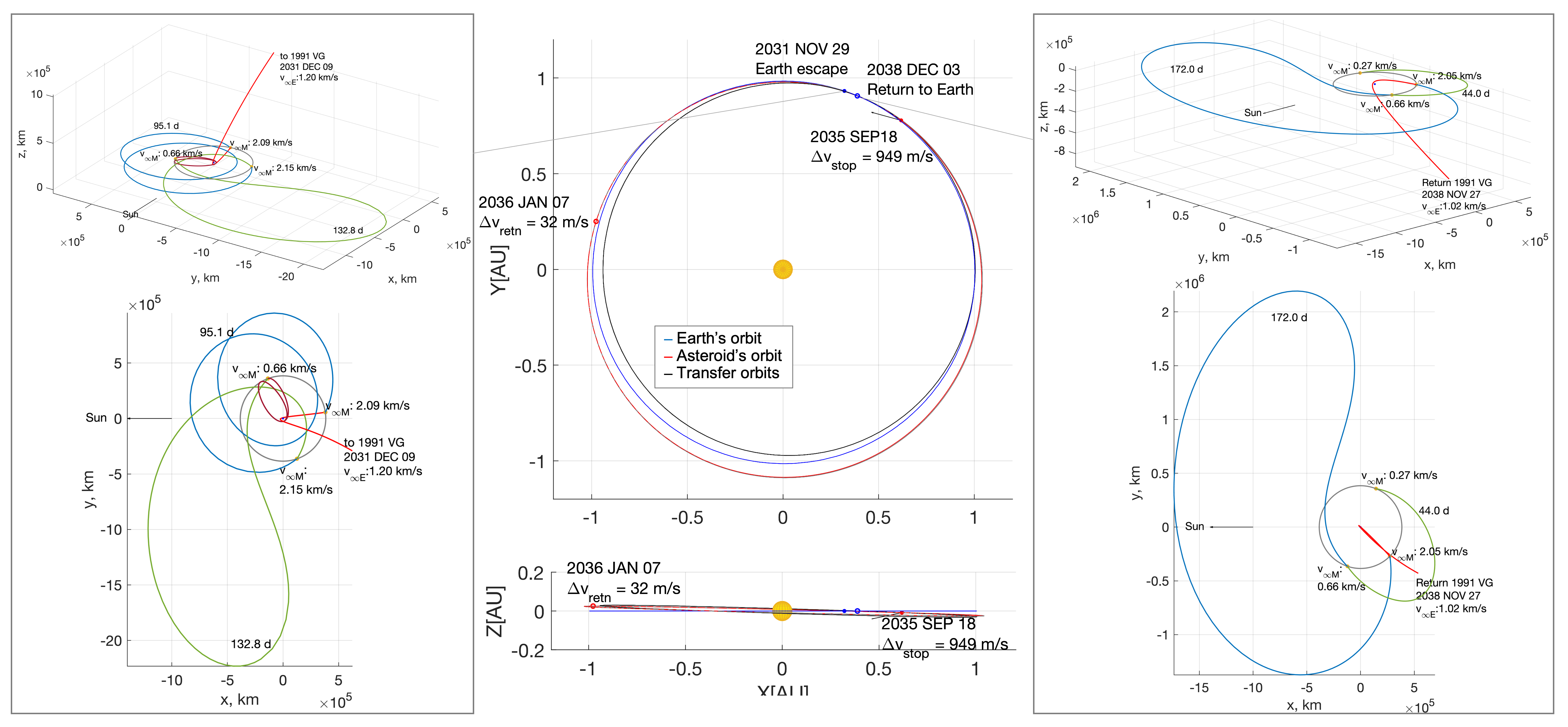}
	\caption{An example mission profile for 1991 VG retrieval, with a gravity-assisted escape phase (left), a heliocentric transfer phase for the asteroid rendezvous and Earth re-encounter (middle), and a gravity-assisted capture phase (right).}
	\label{fig:profile1}
\end{figure}

\begin{figure}[h!t]
	\centering\includegraphics[width=\textwidth]{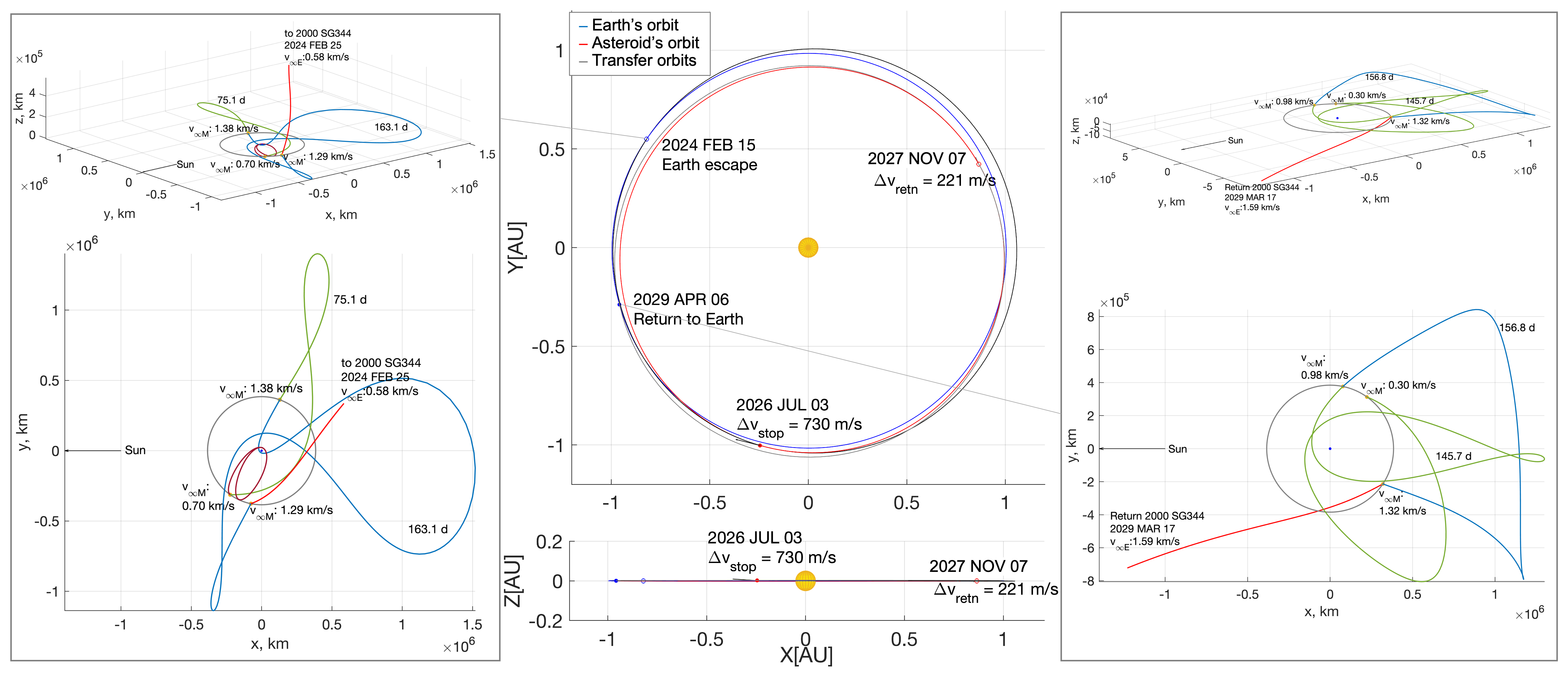}
	\caption{An example mission profile for 2000 SG344 sample return, with a gravity-assisted escape phase (left), a heliocentric transfer phase for the asteroid rendezvous and Earth re-encounter (middle), and a gravity-assisted capture phase (right).}
	\label{fig:profile2}
\end{figure}

If a low-cost asteroid rendezvous is also required, Table~\ref{tab:RetrievalSummary} presents representative low-cost retrieval missions for 16 asteroids, prioritizing the total flight time $ToF_{\text{tot}}$ in the heliocentric phase. To exemplify, Fig.~\ref{fig:profile1} presents a set of trajectory options for the 1991 VG retrieval mission, and Fig.~\ref{fig:profile2} presents the trajectory options for a 2000 SG344 sample-return mission. 

\begin{table}[h!]
	\centering
	\caption{Selected asteroid sample-return and retrieval missions constrained by gravity-assisted asteroid rendezvous and prioritizing the total flight time}\label{tab:RetrievalSummary}
	\begin{adjustbox}{max width=1.6\textwidth,center}
		\begin{tabular}{lllrlrlr}\toprule
			Asteroid & Escape from Earth & Arrival at asteroid & $\Delta v_{\text{stop}}$ [m/s] & Departure for Earth & $\Delta v_{\text{retn}}$ [m/s] & Return to Earth & $ToF_{\text{tot}}$ [day] \\ \midrule
			1991 VG    & 2031 NOV 29 & 2035 SEP 18 & 949 & 2036 JAN 07 & 31.6  & 2038 DEC 03 & 2561 \\
			2000 SG344 & 2024 FEB 15 & 2026 JUL 03 & 730 & 2027 NOV 07 & 221.2 & 2029 APR 06 & 1877 \\
			2006 RH120 & 2024 JAN 27 & 2026 OCT 11 & 828 & 2027 JUL 08 & 438.6 & 2028 OCT 23 & 1731 \\
			2007 UN12  & 2030 FEB 14 & 2033 AUG 22 & 973 & 2033 SEP 26 & 149.0 & 2035 OCT 19 & 2073 \\
			2008 HU4   & 2024 FEB 15 & 2027 JUN 22 & 978 & 2033 FEB 17 & 335.6 & 2036 APR 28 & 4456 \\
			2008 UA202 & 2024 FEB 15 & 2026 AUG 04 & 918 & 2026 DEC 21 & 329.2 & 2028 OCT 05 & 1694 \\
			2009 BD    & 2024 MAY 16 & 2026 OCT 14 & 712 & 2029 JAN 03 & 418.2 & 2033 MAY 23 & 3294 \\
			2010 UJ    & 2024 OCT 20 & 2028 JAN 18 & 638 & 2029 JUL 29 & 410.2 & 2032 DEC 04 & 2967 \\
			2010 UE51  & 2024 NOV 15 & 2027 APR 04 & 784 & 2031 SEP 04 & 470.4 & 2035 NOV 27 & 4029 \\
			2010 VQ98  & 2029 NOV 14 & 2033 JUL 20 & 961 & 2034 SEP 26 & 471.9 & 2037 OCT 29 & 2905 \\
			2011 UD21  & 2031 APR 17 & 2035 APR 21 & 884 & 2035 NOV 17 & 398.2 & 2038 OCT 15 & 2738 \\
			2014 WU200 & 2031 JUN 18 & 2035 JAN 03 & 942 & 2035 JUN 22 & 104.0 & 2038 DEC 18 & 2740 \\
			2012 TF79  & 2024 FEB 15 & 2026 DEC 03 & 839 & 2039 JAN 27 & 397.9 & 2042 OCT 18 & 6820 \\
			2014 WX202 & 2027 MAY 17 & 2030 DEC 07 & 991 & 2032 SEP 27 & 273.9 & 2033 NOV 23 & 2382 \\
			2015 VC2   & 2025 MAR 27 & 2029 JAN 14 & 807 & 2037 NOV 01 & 473.5 & 2041 DEC 31 & 6123 \\
			2016 TB18  & 2024 MAR 30 & 2028 APR 09 & 570 & 2034 DEC 05 & 273.1 & 2036 MAR 24 & 4377
			\\ \bottomrule
		\end{tabular}
	\end{adjustbox}
\end{table}

\section{Discussions}\label{Sec:discussions}
\subsection{Considerations and applications of transfer options}
\begin{figure}[h!t]
	\begin{center}
		\begin{tabular}{c:c}
			\includegraphics[width=0.3\textwidth]{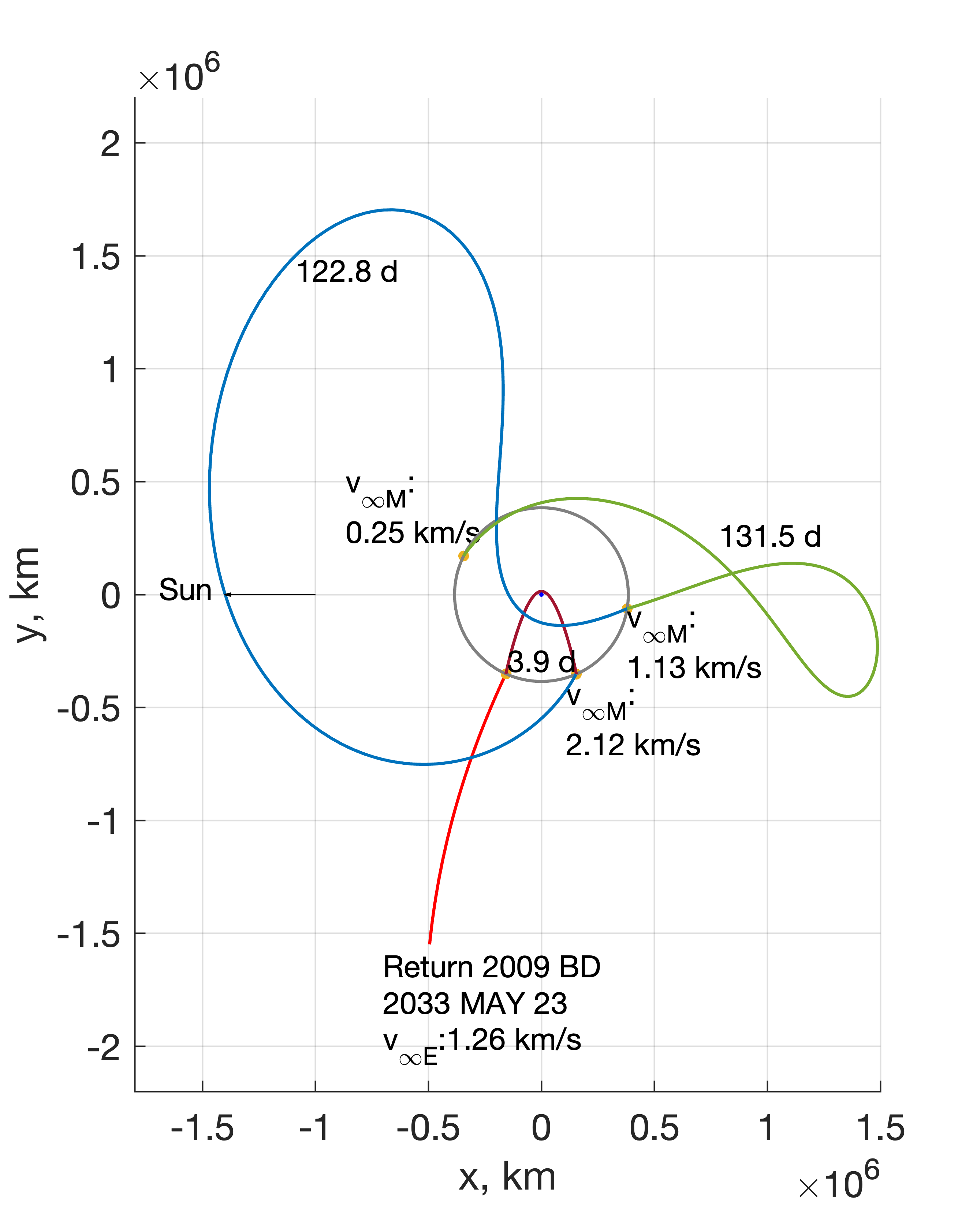} &
			\includegraphics[width=0.3\textwidth]{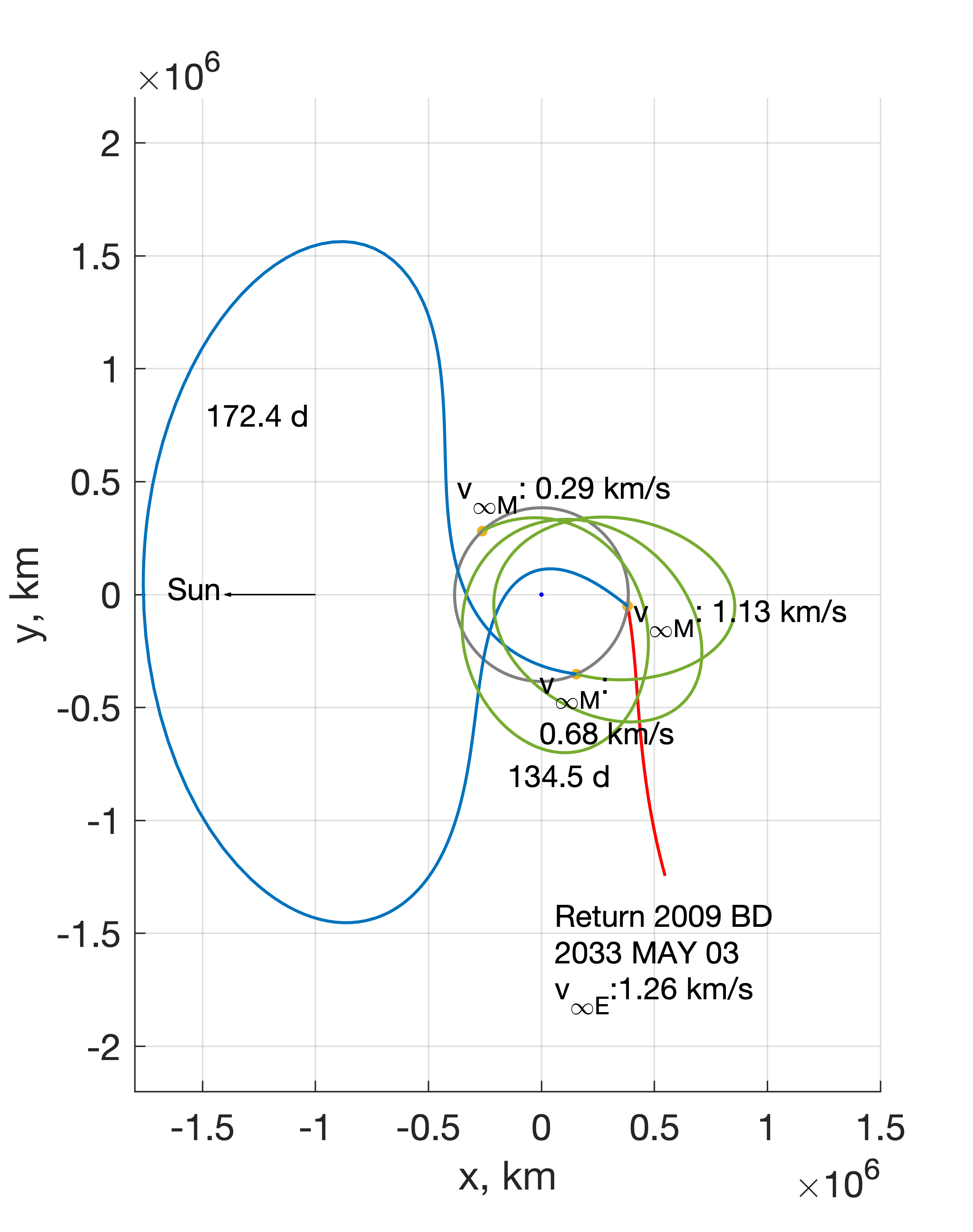} \\\hdashline 
			\includegraphics[width=0.3\textwidth]{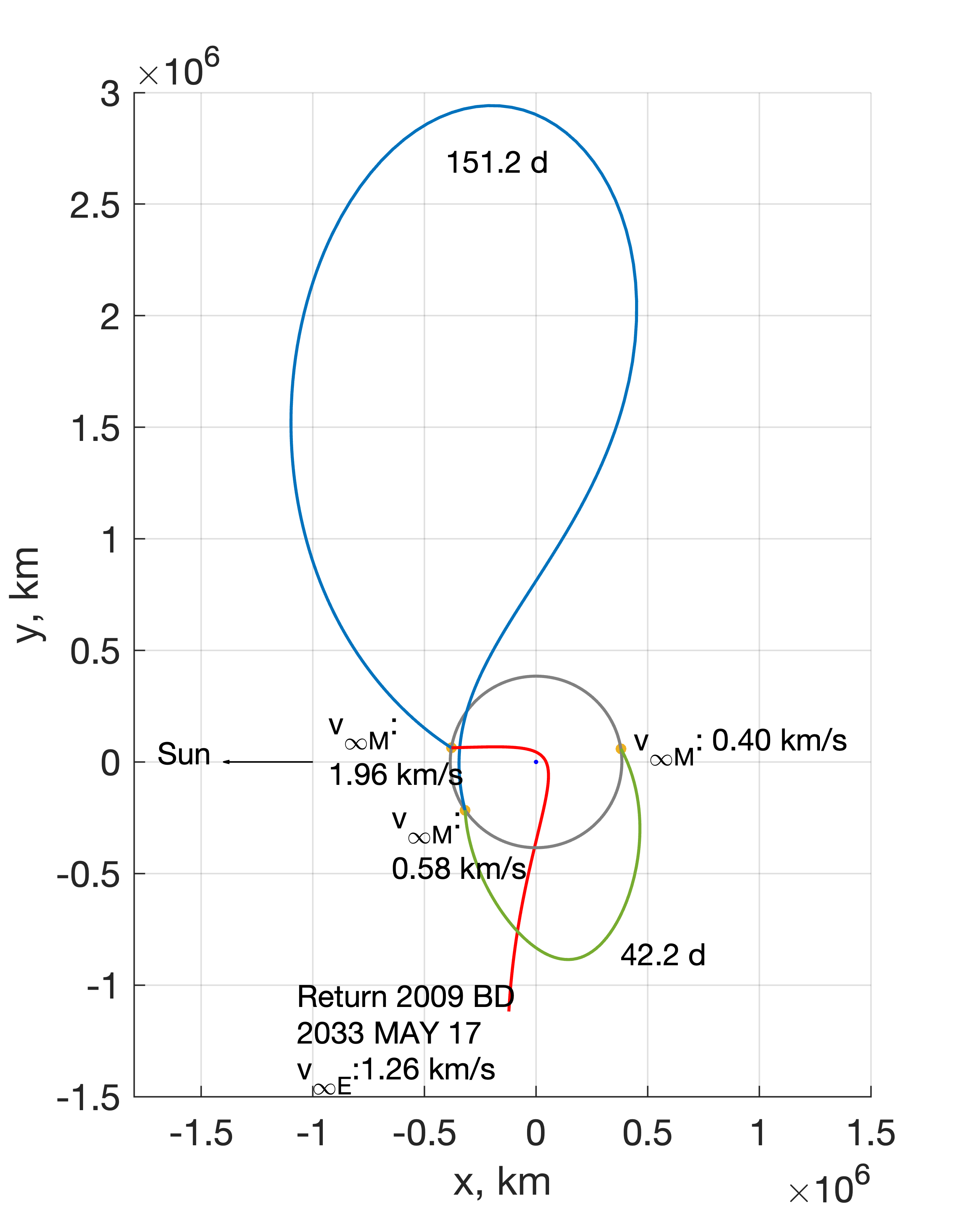} &\includegraphics[width=0.3\textwidth]{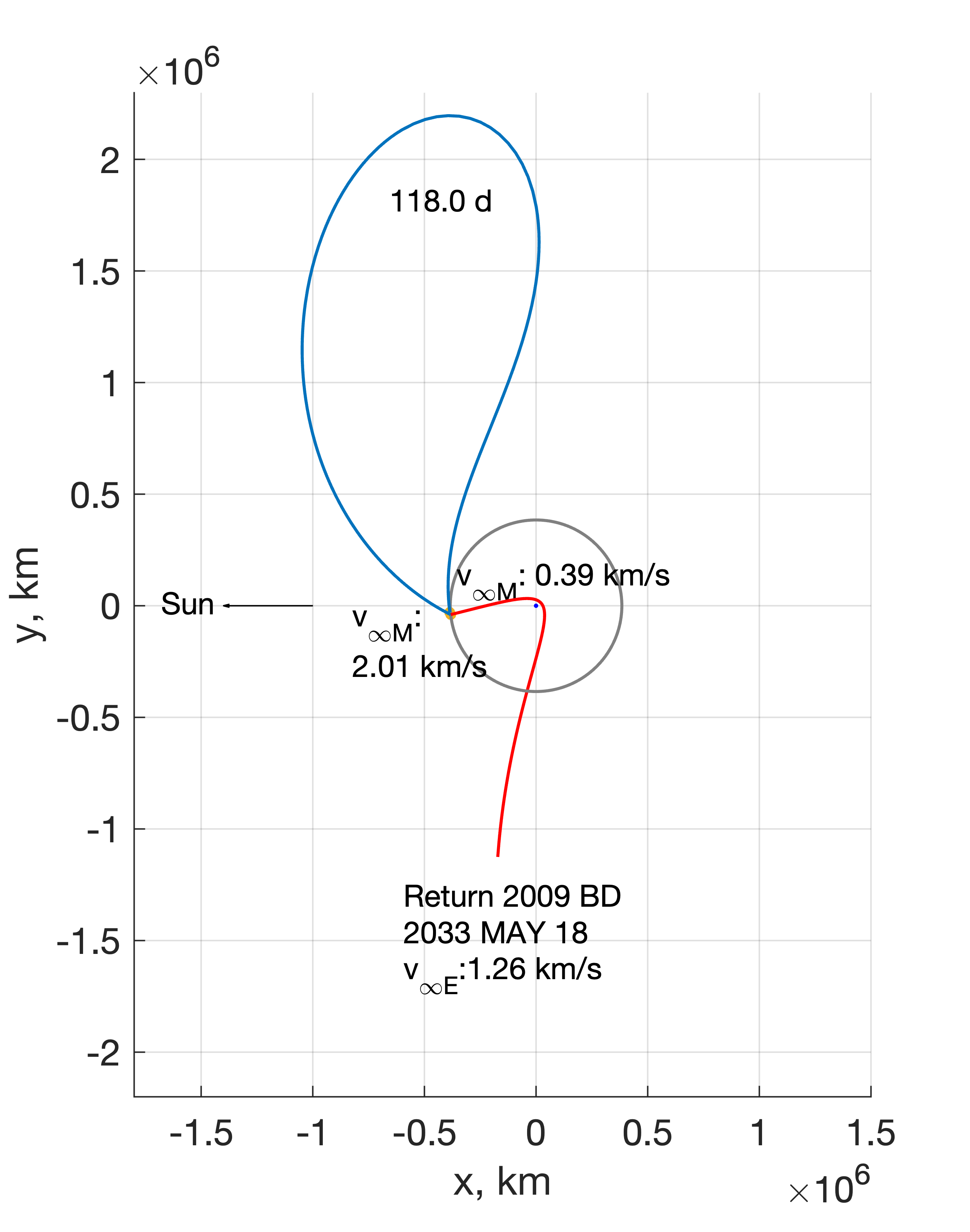}
		\end{tabular} 
		\caption{Available options of lunar swingby sequence for capturing the 2009 BD sample arriving around 2033 May 15  (synodic frame).}
		\label{fig:demo-traj}  
	\end{center}
\end{figure}

Figure \ref{fig:demo-traj} presents options of gravity-assisted trajectories for capturing the sample from 2009 BD arriving in the vicinity of the Earth around 2033 May 15. The upper-left panel demonstrates the capture trajectory adopting a short inner M-M transfer and two Sun-perturbed transfers. The upper-right panel demonstrates another option adopting a three-rev transfer. The inner M-M transfer is not always necessary. Furthermore, it is considered not very practical for the challenging GNC operation to be performed within the short period between two consecutive lunar swingbys. As for multi-rev M-M transfers, as they regularly pass the orbit of the Moon, they can be perturbed by the gravity of the Moon, which is not described by the ``simplified" model used in this work. Hence, the designed trajectory involving multi-rev M-M transfers cannot be used directly without further refinement in the high-fidelity model. Nevertheless, this simplified model provides insights into orbit dynamics and handy approaches to orbit design, which are tricky to achieve in a high-fidelity model. A number of transfer options generated by this method can serve as the initial guesses for orbit optimization in a high-fidelity model. The trajectory refinement and optimization may result in a few small $\Delta v$ during the courses of the M-M transfers, which is to either avoid or to make use of the gravity of the Moon. Lantoine and McElrath (2014)~\cite{Lantoine2014} have demonstrated that the Sun-driven lunar swingby sequence designed in this simplified model can be successfully transferred to the ephemeris model without significant correction. 

In addition, options without a short inner transfer or multi-rev transfer may be available. The bottom two panels present the trajectory options for different first lunar phase angles (or the Earth encounter dates). In the bottom-left panel, the retrograde transfer arc is greatly altered around the apogee in the 2nd quadrant, and the orbit finally becomes a posigrade orbit with a small $v_{\infty\text{M}}$. However, the two SPMTs take as long as 200 days. The bottom-right panel demonstrates a more favorable option, where only one large M-M transfer arc and one lunar swingby are needed, which results in a minimum flight time (i.e., 120 days) and GNC challenge. To summarize, it is advantageous to have a complete database of SPMT (e.g., including multi-rev transfers) and fully explore all possible sequences, so as to gain flexibility and robustness with backup options.

\subsection{Three-dimensional Moon-to-Moon transfers}\label{sec:3Ddiscussion}
\begin{figure}[h!t]
	\begin{center}
		\includegraphics[width=1\hsize]{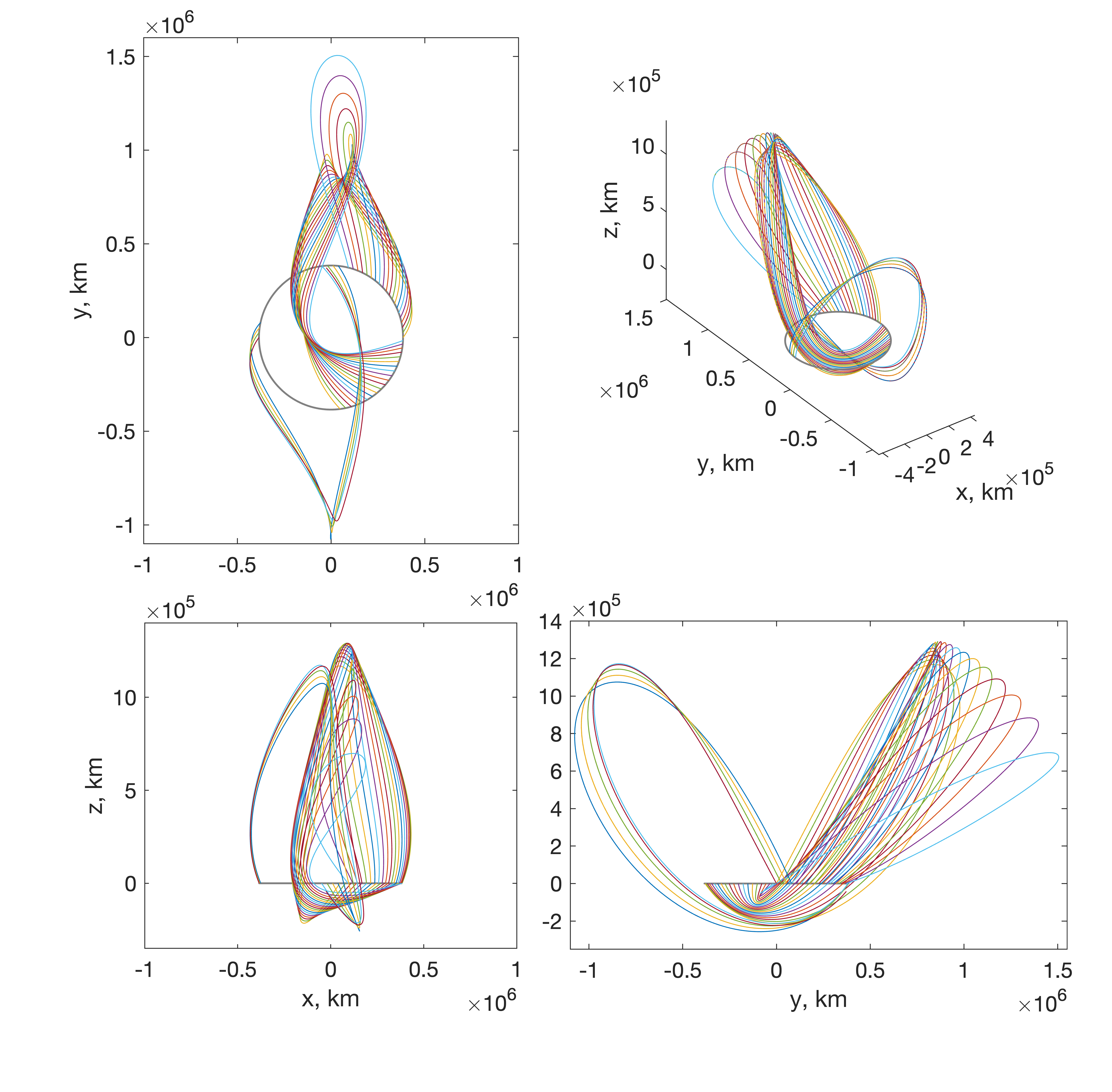}
		\caption{Example three-dimensional Sun perturbed Moon-to-Moon transfers  (synodic frame).}
		\label{fig-SPMT3D}
	\end{center}
\end{figure}

The three-dimensional (3D) SPMT also exists and can serve to connect two lunar encounters. A database of 3D transfers has also been computed using a similar routine to that presented in Sec.~\ref{sec:database}. Example 3D trajectories (i.e., \(v_{\infty\text{M}} = 1.2~\si{\kilo\meter/\second}\); ``oo 1:2'' family) are shown in Fig.\ref{fig-SPMT3D}. Observation of the 3D solutions shows that $v_{\infty\text{M}}$ at the lunar encounters are generally large (i.e., $>$ 0.8 km/s). That is because the out-of-plane component of $\bf{v}_{\infty\text{{M}}}$ cannot be significantly varied by solar tides. A 3D transfer can change the connecting phase, similar to what a non-perturbed lunar resonant orbit can do, and thus add more patching options. Nevertheless, this is considered not efficient especially for the considered problem, in which $ToF$ and segments of perturbed M-M transfers are limited.  
Therefore, for the escape phase, it is preferred to perform a sequence of planar M-M transfers and lunar swingbys before the final bend to the out-of-plane trajectory heading for the asteroid. Similarly, for the capture phase, it is preferred to first bend the out-of-plane incident trajectory onto the plane for the following effective gravity assists performed in the plane. 

\section{Conclusions}
The capacity and application of the Sun-driven lunar swingby sequence are presented in this paper. A variety of types of Moon-to-Moon transfers, namely, the short transfer, Sun-perturbed transfer, multi-rev transfer, and three-dimensional transfer, were discussed. Analyses with the ``Swingby-Jacobi'' graph indicate that 
\begin{enumerate}[resume]
	\item A Sun-driven lunar swingby sequence can access a range of the Jacobi integral from -3.0009 to -2.9946. This range encloses 657 potential asteroids currently cataloged, which can reduce the effort of target search and trajectory optimization.
	\item $C_3$ for escape can be increased to 4.8 km$^2$/s$^2$. Reversely, objects of this $C_3$ level can possibly (depending on the direction) be captured.  In particular, this technique can at least reach or absorb a $v_{\infty\text{E}}$ of 1.46 km/s (i.e., $C_3$ of 2.13 km$^2$/s$^2$) in all directions. 
\end{enumerate}

Heliocentric transfer trajectories between the Earth and potential candidates were optimized, and massive sequences of lunar swingbys were explored thanks to the database of Sun-perturbed Moon-to-Moon transfers. Results show that Sun-driven lunar swingby sequences can enable:
\begin{enumerate}[resume]
	\item spacecraft to rendezvous with 48 asteroids by 2043, at a $\Delta v$ cost of less than 1 km/s after the first lunar encounter. Even without the rendezvous $\Delta v$, spacecraft can still flyby and observe the asteroids such as 1991 VG, 2000 SG344, 2014 YD, 2016 TB18, and 2017 BN93 for more than one week. This is still significant for a low-cost mission. 
	\item 25 asteroid samples to be captured by 2043, at a $\Delta v$ cost of less than 500 m/s for the Earth re-encounter (while lunar orbit injection requires an additional 20 to 350 m/s of $\Delta v$). In particular, given the capabilities of state-of-the-art launch and propulsion systems, retrieving the entire 10-m asteroid 1991 VG in 2038, which requires a total impulse of $\SI{4.6e4}{\kilo\newton\cdot\second}$,  is considered feasible. 
\end{enumerate}

\appendix
\subsection*{Appendix (Patching the heliocentric phase and the lunar swingby sequence)}

\begin{figure}[h!t]
	\begin{center}
		\includegraphics[width=0.6\hsize]{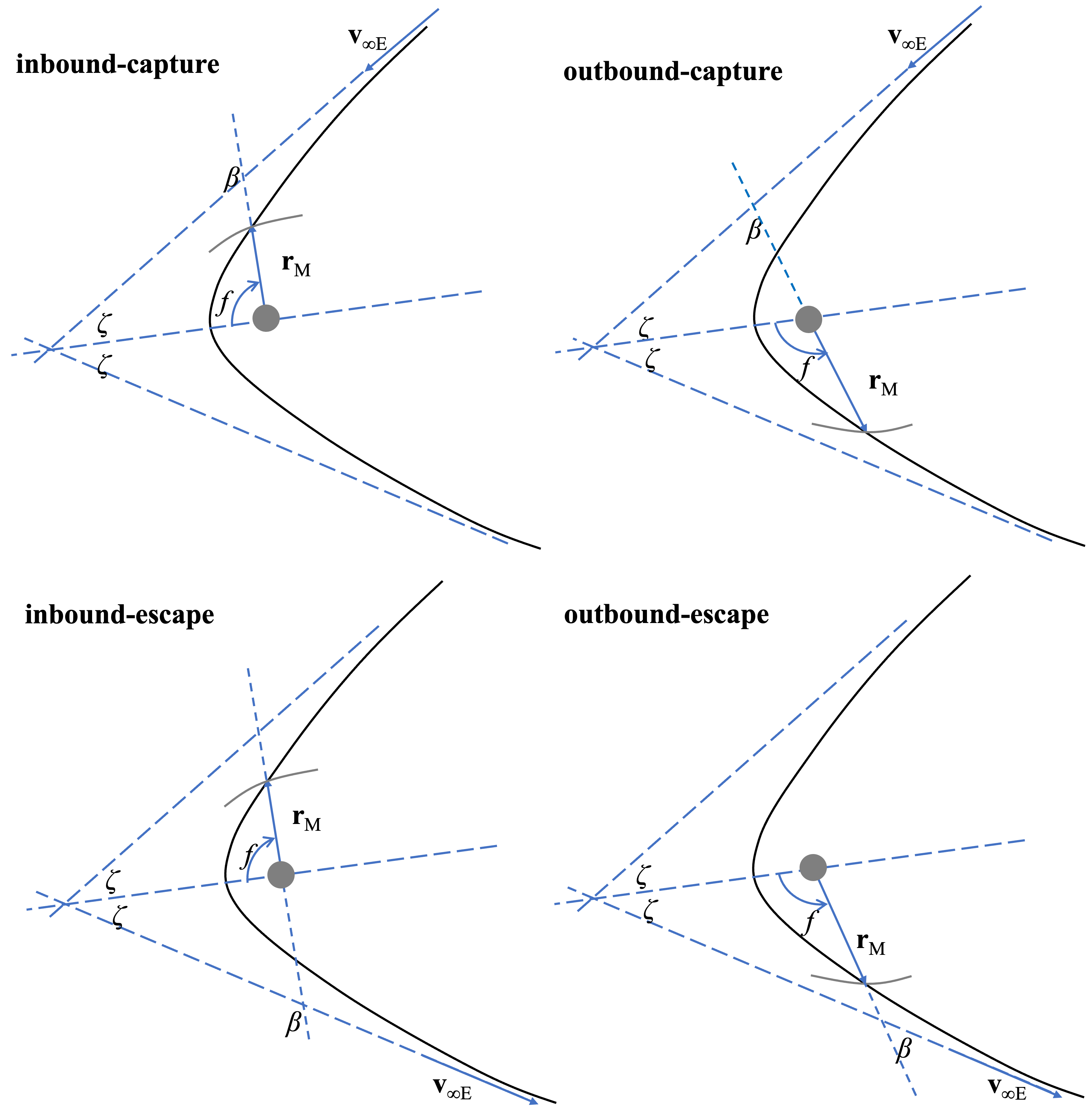}
		\caption{Situations of lunar encounters with a hyperbolic orbit on the inbound leg and outbound leg.}
		\label{fig-encounter}
	\end{center}
\end{figure}

Considering the capture case, given the incident $\bf{v}_{\infty\text{E}}$ and the Moon’s position vector $\bf{r}_\text{M}$ as the encounter conditions, the orbital elements of the hyperbolic orbit about Earth can be computed with the routine presented in this appendix. There are two situations of encounters. One kind of encounter takes place on the inbound leg, and another kind of encounter takes place on the outbound leg (see Fig.\ref{fig-encounter}). 
For the inbound-capture case, the relationship between the true anomaly $f$ at the encounter, turn angle $\zeta$ (i.e., half the bending angle), and the angle $\beta$ between $\bf{v}_{\infty\text{E}}$ and $\bf{r}_\text{M}$ is expressed by:
\begin{equation}
	\cos f=\cos (\beta -\zeta )
\end{equation}
which can be rewritten as:
\begin{equation}
	\cos f=\sin \beta \sin \zeta +\cos \beta \cos \zeta 
\end{equation}
$\zeta$ is related to the eccentricity $e_\text{p}$ of the orbit about Earth. The subscript ``p'' indicates the orbit about the planet. Hence, 
\begin{equation}
	\cos f=\frac{\sqrt{{{e}_\text{p}^{2}}-1}}{e_\text{p}}\sin \beta +\frac{1}{e_\text{p}}\cos \beta 
\end{equation}
At the encounter, the radius of the hyperbolic orbit is the same as the radius of Moon's orbit, which is expressed as:
\begin{equation}
	{{r}_\text{M}}=\frac{a_\text{p}(1-{{e}_\text{p}^{2}})}{1+e_\text{p}\cos f}
\end{equation}
where the semi-major axis $a_\text{p}$ is related to ${v}_{\infty\text{E}}$ through,
\begin{equation}
	a_\text{p}=-G_{\text{E}} /v_{\infty \text{E}}^{2}
\end{equation}
where $G_{\text{E}}$ is the gravitational parameter of the Earth. Integrating the last two equations yields a quadratic function of \(\sqrt{{{e}_\text{p}^{2}}-1}\),
\begin{equation}
	a_\text{p}({e}_\text{p}^{2}-1)+{{r}_\text{M}}\sin \beta \sqrt{{{e}_\text{p}^{2}}-1}+{{r}_\text{M}}(1+\cos \beta )=0
\end{equation}
whose solution is computed from,
\begin{equation}
	\sqrt{{{e}_\text{p}^{2}}-1}=\frac{-{{r}_\text{M}}\sin \beta -\sqrt{{r}_\text{M}^{2}{{\sin }^{2}}\beta -4a_\text{p}{{r}_\text{M}}(1+\cos \beta )}}{2a_\text{p}}
\end{equation}
With $e_\text{p}$ known, the orbit can be determined. The true anomaly is computed from:
\begin{equation}
	f=-\arccos \left[\frac{a_\text{p}(1-{{e}_\text{p}^{2}})-{{r}_{\text{M}}}}{{{r}_{\text{M}}}e_\text{p}}\right]
\end{equation}
Other desired orbital information at the lunar encounter can be derived. The steps of deriving $e_\text{p}$ and $f$ for the outbound-capture, inbound-escape, and outbound-capture cases are similar. The corresponding expressions are directly given as follows.
For the outbound-capture case:
\begin{align}
	\sqrt{{{e}_\text{p}^{2}}-1}&=\frac{{{r}_\text{M}}\sin \beta -\sqrt{{r}_\text{M}^{2}{{\sin }^{2}}\beta -4a_\text{p}{{r}_\text{M}}(1+\cos \beta )}}{2a_\text{p}} \\
	f&=\arccos \left[\frac{a_\text{p}(1-{{e}_\text{p}^{2}})-{{r}_{\text{M}}}}{{{r}_{\text{M}}}e_\text{p}}\right]
\end{align}
For the inbound-escape case:
\begin{align}
	\sqrt{{{e}_\text{p}^{2}}-1}&=\frac{{{r}_\text{M}}\sin \beta -\sqrt{{r}_\text{M}^{2}{{\sin }^{2}}\beta -4a_\text{p}{{r}_\text{M}}(1-\cos \beta )}}{2a_\text{p}} \\
	f&=-\arccos  \left[\frac{a_\text{p}(1-{{e}_\text{p}^{2}})-{{r}_{\text{M}}}}{{{r}_{\text{M}}}e_\text{p}}\right]
\end{align}
For the outbound-escape case:
\begin{align}
	\sqrt{{{e}_\text{p}^{2}}-1}&=-\frac{{{r}_\text{M}}\sin \beta +\sqrt{{r}_\text{M}^{2}{{\sin }^{2}}\beta -4a_\text{p}{{r}_\text{M}}(1-\cos \beta )}}{2a_\text{p}} \\
	f&=\arccos  \left[\frac{a_\text{p}(1-{{e}_\text{p}^{2}})-{{r}_{\text{M}}}}{{{r}_{\text{M}}}e_\text{p}}\right]
\end{align}

\subsection*{Acknowledgements}
\label{acknowledgements}

This work was primarily conducted at the Technology and Engineering Center for Space Utilization, Chinese Academy of Sciences. The massive exploration of sequences of Moon-to-Moon transfers and lunar swingbys benefited from the computer cluster \texttt{cerfeuil}, financed and managed by IMCCE/Paris Observatory. The author sincerely appreciates the valuable comments from the anonymous reviewers. 

\subsection*{Declaration of competing interest}

The authors have no competing interests to declare that are relevant to
the content of this article.

\section*{References}

\bibliographystyle{astrobib}
\bibliography{refs}

\subsection*{Author biography}

\begin{biography}[photo_HC]{Hongru Chen} is an Assistant Professor at Kyushu University, Japan. She received her Bachelor's degree from Northwestern Polytechnical University, China, in 2010, and her Ph.D. from Kyushu University, in 2015, both in aerospace engineering. She did her Ph.D. thesis at ISAS/JAXA, once worked at the Chinese Academy of Sciences and IMCCE/Paris Observatory, and participated in JAXA and CNES planetary projects, such as PROCYON, DESTINY, and MMX. Her research interests include astrodynamics, orbit design, nano-satellite engineering, and atmosphere modeling. 
\end{biography}


\vspace*{2.6em}
\subsection*{Graphical table of contents}

\begin{figure*}[h!t]
	\begin{center}
		\includegraphics[width=\hsize]{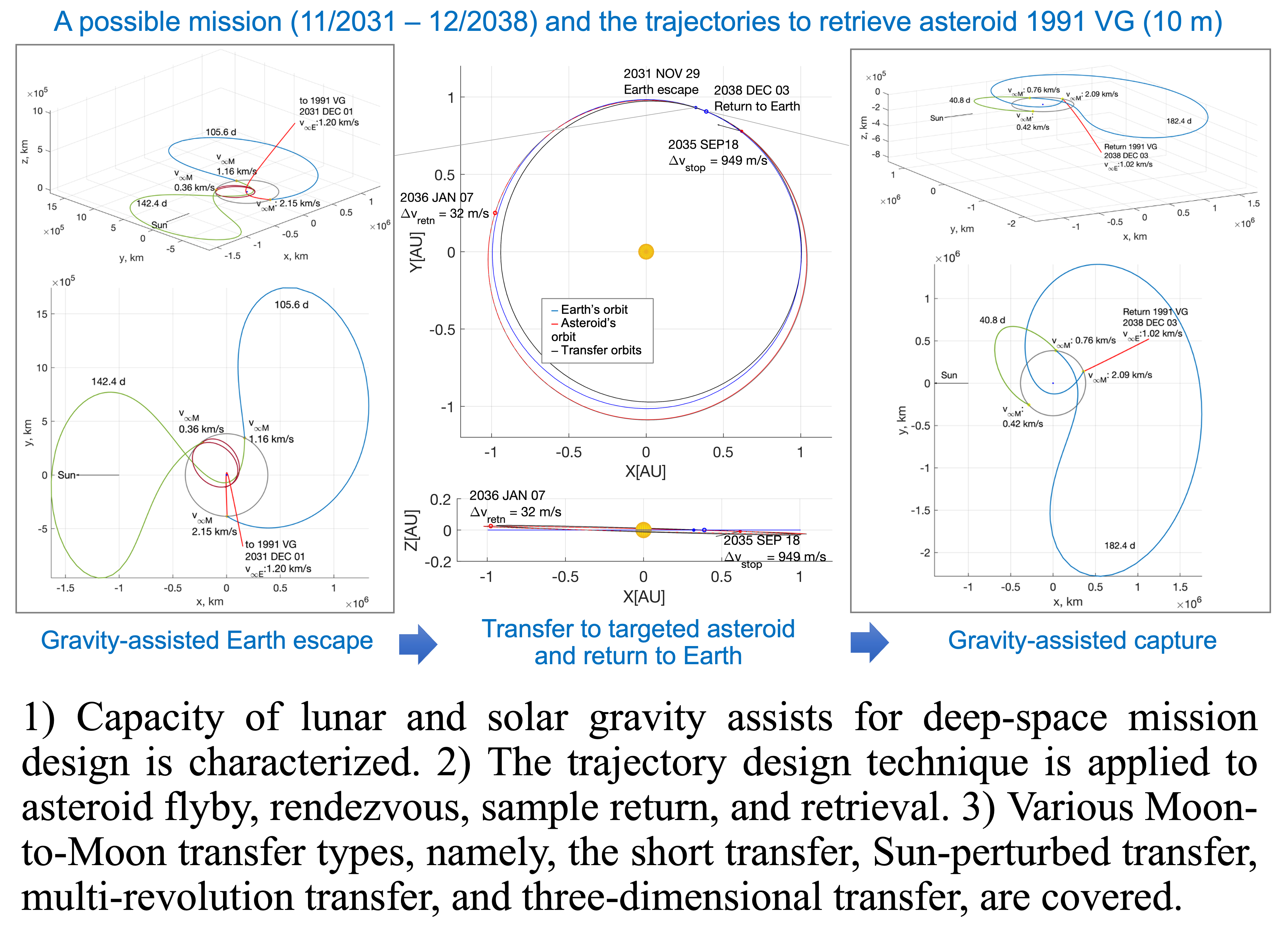}
		\caption{Graphical abstract of the paper}
	\end{center}
\end{figure*}

\end{document}